\documentclass[12pt,a4paper,onecolumn,oneside,abstract=true]{scrartcl}

\usepackage[utf8]{inputenc}
\usepackage[main=english,ngerman]{babel}
\usepackage{lmodern}
\usepackage[T1]{fontenc}
\usepackage{amssymb,amsmath,amsthm,mathtools}
\usepackage{amsfonts}
\usepackage{lipsum}
\usepackage{multirow}
\usepackage{comment}
\usepackage{booktabs}
\usepackage{graphics,graphicx}
\usepackage[dvipsnames]{xcolor}
\usepackage{dsfont}
\usepackage[mathscr]{euscript}
\usepackage{enumerate}							
\usepackage[]{listofsymbols}
\usepackage[round]{natbib}
\usepackage{authblk}
\usepackage{etoolbox} 
\usepackage{caption} 
\usepackage[flushleft]{threeparttable} 
\usepackage{bm} 
\usepackage{algorithm} 
\usepackage{algpseudocode} 
\usepackage{gensymb} 

\DeclareUnicodeCharacter{00A0}{ }

\newcommand{\indicatorset}[1]{\mathds{1}_{#1}}

\newcommand{\toinucp}{\xrightarrow{{\lower2pt\hbox{$\!\mathrm{\scriptstyle ucp}\!$}}}}

\newcommand\sbullet[1][.5]{\mathbin{\vcenter{\hbox{\scalebox{#1}{$\bullet$}}}}}

\opensymdef
\newsym[Set of all natural numbers excluding zero]{na}{\mathbb{N}}
\newsym[Set of all natural numbers including zero]{nawz}{\na_{0}}
\newsym[Set of all real numbers]{re}{\mathbb{R}}
\newsym[Set of all nonnegative real numbers]{replus}{\re_{+}}
\newsym[Base $\sigma$-field]{Fcal}{\mathcal{F}}
\newsym[Base filtration]{FF}{\mathbb{F}}
\newsym[Base probability measure]{PP}{\mathbb{P}}
\newsym[Alternative probability measure]{QQ}{\mathbb{Q}}
\newsym[Borel $\sigma$-field]{B}{\mathcal{B}}
\newsym[Topology]{T}{\mathcal{T}}
\newsym[Space of all real valued semimartingales]{SM}{\mathscr{S}}
\newsym[Space of all real valued caglad processes]{LD}{\mathscr{L}}
\newsym[Space of all real valued cadlag processes]{DD}{\mathscr{D}}
\newsym[Set of controls]{MM}{\mathbb{M}}
\newsym[Set of controlled processes]{CP}{\mathcal{NN}(\Theta, \psi)}
\newsym[Process that generates the (augmented) base filtration]{Y}{Y}
\newsym[Integrands, trading strategies]{X}{X}
\newsym[Semimartingale integrators]{SI}{S}
\closesymdef

\theoremstyle{plain}
\newtheorem{theorem}{Theorem}[section]

\theoremstyle{definition}

\theoremstyle{remark}
\newtheorem{remark}[theorem]{Remark}

\DeclareMathOperator*{\argmin}{arg\,min} 
\DeclareMathOperator*{\argsup}{arg\,sup}

\makeatletter

\DeclareOldFontCommand{\bf}{\normalfont\bfseries}{\mathbf}

\newcommand{\makeoverbar}[7]{%
\setbox0=\hbox{$\m@th#2\mkern#5mu{{}#3{}}\mkern#6mu$}%
\setbox1=\null \dimen@=#4\fontdimen8#13 \dimen@=3.5\dimen@
\advance\dimen@ by \ht0 \dimen@=-#7\dimen@ \advance\dimen@ by \wd0
\ht1=\ht0 \dp1=\dp0 \wd1=\dimen@
\dimen@=\fontdimen8#13 \fontdimen8#13=#4\fontdimen8#13
\rlap{\hbox to \wd0{$\m@th\hss#2{\overline{\box1}}\mkern#5mu$}}
\fontdimen8#13=\dimen@}

\def\makeunderbar#1#2#3#4#5#6#7{%
\setbox0=\hbox{$\m@th#2\mkern#5mu{{}#3{}}\mkern#6mu$}%
\setbox1=\null \dimen@=#4\fontdimen8#13 \dimen@=3.5\dimen@
\advance\dimen@ by \dp0 \dimen@=-#7\dimen@ \advance\dimen@ by \wd0
\ht1=\ht0 \dp1=\dp0 \wd1=\dimen@
\dimen@=\fontdimen8#13 \fontdimen8#13=#4\fontdimen8#13
\rlap{\hbox to \wd0{$\m@th\hss#2{\underline{\box1}}\mkern#5mu$}}
\fontdimen8#13=\dimen@}

\patchcmd{\@maketitle}
  {{\usekomafont{date}{\@date \par}}%
    \vskip \z@ \@plus 1em}
  {}
  {}{}

\def\thanks#1{\protected@xdef\@thanks{\@thanks
        \protect\footnotetext{#1}}}

\makeatother

\allowdisplaybreaks

\usepackage[pagebackref=false]{hyperref}
\hypersetup{colorlinks=true,linkcolor=blue,citecolor=blue,filecolor=blue,urlcolor=blue,pdfauthor={Aleksandar Arandjelovic}}
\hypersetup{final}

\setkomafont{disposition}{\bfseries}
\usepackage[top=2.5cm, bottom=2.5cm, left=1cm, right=1cm]{geometry} 
\usepackage[onehalfspacing]{setspace} 
\usepackage{parskip}

\title{Solving stochastic climate-economy models: A deep least-squares Monte Carlo approach}

\author[1,2]{Aleksandar~Arandjelovi\'{c}}
\author[3]{Pavel~V.~Shevchenko%
\thanks{Date: \today \\ \hspace*{-0.85em} Corresponding author: \href{mailto:pavel.shevchenko@mq.edu.au}{pavel.shevchenko@mq.edu.au}}}
\author[4]{Tomoko~Matsui}
\author[5]{Daisuke~Murakami}
\author[6]{Tor~A.~Myrvoll}

\affil[1]{Department of Mathematics, ETH Zurich, Switzerland}
\affil[2]{Institute for Statistics and Mathematics, Vienna University of Economics and Business, Austria}
\affil[3]{Department of Actuarial Studies and Business Analytics, Macquarie University, Australia}
\affil[4]{Department of Statistical Modeling, Institute of Statistical Mathematics, Japan}
\affil[5]{Department of Statistical Data Science, Institute of Statistical Mathematics, Japan}
\affil[6]{Department of Electronic Systems, Norwegian University of Science and Technology, Norway}

\begin{document}

\maketitle
\thispagestyle{empty}

\vspace{-1cm}

\abstract{
\noindent
Stochastic versions of recursive integrated climate-economy assessment models are essential for studying and quantifying policy decisions under uncertainty.
{\color{black}However, as the number of state variables  and stochastic shocks increases, solving these models via deterministic
grid-based dynamic programming (e.g., value-function iteration / projection on a discretized grid
over continuous state variables, typically coupled with discretized shocks) becomes computationally
infeasible, and simulation-based methods are needed.}
The least-squares Monte Carlo (LSMC) method has become popular for solving optimal stochastic control problems in quantitative finance.
In this paper, we extend the application of the LSMC method to stochastic climate-economy models.
{\color{black}We exemplify this approach using a stochastic version of the DICE model with five key uncertainty sources highlighted in the literature.}
To address the complexity and high dimensionality of these models, we incorporate deep neural network approximations in place of standard regression techniques within the LSMC framework.
Our results demonstrate that the deep LSMC method can be used to efficiently derive optimal policies for climate-economy models in the presence of uncertainty.

}

\clearpage

\setlength{\parindent}{0pt}
\setlength{\parskip}{.5\baselineskip plus 2pt}

\section{Introduction}\label{introduction}
The analysis of climate-economy policies is typically performed using Integrated Assessment Models (IAMs) that describe the complex interplay between the climate and the economy via deterministic equations.
In order to account for stochastic shocks when finding optimal mitigation policies adapted to climate and economic variables that are evolving stochastically over time, a recursive dynamic programming implementation of integrated assessment models is required.
This is a significantly harder computational problem to solve compared to the deterministic case.
Seminal contributions to solving IAMs as optimal decision-making problems in the presence of uncertainty include \cite{kelly1999}, \cite{kelly2001solving}, \cite{leach2007}, \cite{traeger2014}, and \cite{cai2019}.
All these studies are based on variants of the so-called dynamic integrated climate-economy (DICE) model extended to include stochastic shocks to the economy and climate.
The DICE model is one of the three main IAMs (the other two being FUND and PAGE) used by the United States government to determine the social cost of carbon; see \cite{scc2016technical}.
It has been regularly revised over the last three decades, with the first version dating back to \cite{nordhaus1992dice}.
It balances parsimony with realism and is well documented with all published model equations; in addition, its code is publicly available, which is an exception rather than the rule for IAMs.
At the same time, it is important to note that IAMs, and the DICE model in particular, have significant limitations (in the model structure and model parameters), which have been criticized and debated in the literature (see the discussions in \cite{ackerman2009limitations,pindyck2017use,grubb2021modeling,weitzman2011fat}).
{\color{black}Recent work has also raised a more specific concern about the \emph{benchmark calibration} of the reduced-form climate emulator used in DICE-2016: \citet{dietz2021climatedynamics} and
\citet{folini2025climate} document that the standard DICE-2016 calibration can deviate from climate-science benchmarks for key aspects of the transient warming response (and related heat-uptake dynamics). In this paper, we retain the DICE-2016R2 calibration as a widely used benchmark to facilitate comparison with the IAM literature and to keep the focus on the numerical method; our quantitative results should therefore be interpreted conditional on this benchmark climate block.}
Despite the criticism, the DICE model has become the iconic typical reference point for climate-economy modeling and is used in our study.

The original deterministic DICE model is solved as a global optimization problem using the General Algebraic Modeling Language (GAMS)\footnote{\url{https://www.gams.com/}}, a high-level programming language for mathematical modeling.
Its stochastic extensions mentioned in the above-mentioned studies require implementations of recursive dynamic programming to find optimal climate policies under uncertainty\footnote{If required, the deterministic DICE model can be solved as a recursive dynamic programming problem too.}.
This is subject to the curse of dimensionality, and these studies are limited to only one or two stochastic variables.
Even in this case, computations take several million core hours on a modern supercomputer (see, for instance, \citet{cai2019}).
Therefore, simulation methods are needed to handle models with many state variables and multiple shocks to reduce the computational burden.

The least-squares Monte Carlo (LSMC) method for solving multi-dimensional stochastic control problems has gained popularity in recent years due to its effectiveness in dealing with high dimensional problems and because it imposes fewer restrictions on the constraints and allows for flexibility in the dynamics of the underlying stochastic processes.
The idea is based on simulating random paths of the underlying stochastic variables over time and replacing the conditional expectation of the value function in the Bellman backward recursive solution of the stochastic control problem with an empirical least-squares regression estimate.
The transition density of the underlying process is not even required to be known in closed form; one just needs to be able to simulate the underlying processes.
The LSMC method was originally developed in \cite{longstaff2001valuing} and \cite{tsitsiklis2001regression}.
The convergence properties of this method are examined in \cite{belomestny2010regression, belomestny2011pricing}, and \cite{aid2014probabilistic}.
The LSMC method was originally developed for pricing American options where the state variables are not affected by the control.
Later, an extension of the LSMC method with control randomisation was developed in \cite{kharroubi2014numerical} to handle endogenous state variables (i.e.\ state variables that are affected by controls).
When applied to stochastic control problems that aim to optimize an expected utility, some further extensions are needed as proposed in \cite{andreasson2022} and \cite{andreasson2024optimal} to achieve a stable and accurate solution.

{\color{black} %
Our numerical backbone builds on the bias-corrected endogenous-state least-squares Monte Carlo recursion for multi-period expected-utility control developed by \citet{andreasson2022}. The present paper does not claim a new Bellman recursion; rather, it contributes the methodological and implementation extensions required to make this framework practically usable for stochastic integrated assessment models. In particular, relative to \citet{andreasson2022} we: (i) couple the bias-corrected recursion with deep neural-network regressors for continuation-value conditional expectations in a moderately high-dimensional climate-economy state; (ii) introduce numerically critical stabilization choices in the stochastic DICE setting (post-decision states, normalization of rapidly growing economic variables, and low-discrepancy sampling); and (iii) demonstrate how a single backward recursion can be reused for uncertainty quantification (e.g.\ distributions of the social cost of carbon and variance-based Sobol' indices) via forward simulation under the resulting optimal feedback policy.}

{\color{black}
Regression approximations of conditional expectations have a long tradition in computational macroeconomics. In particular, the \emph{parameterized expectations} approach of \citet{den1990solving} computes expectations appearing in equilibrium conditions (typically Euler equations) via forward in time simulation with regression; \citet{duffy2001approximating} extends this idea using neural networks, and \citet{valaitis2024machine} revisits related projection ideas from a modern machine-learning perspective in macro-finance models. Our method is different: backward in time, we approximate the continuation value conditional expectation in the \emph{Bellman recursion} using least-squares Monte Carlo with control randomization and deep neural network regressors, and then recover the optimal control by a separate
numerical maximization step. This makes the method directly applicable to controlled, high-dimensional climate-economy dynamics with multiple shocks and policy constraints, while keeping the learning step (conditional expectation) and the optimization step (policy choice) cleanly separated.
}

In this paper, we demonstrate how the LSMC method can be adapted to solve the recursive dynamic programming problem of stochastic IAMs.
{\color{black}Our goal is not to deliver a new DICE specification or to claim state-of-the-art climate-economic calibration; rather, we use the standard DICE-2016R2 setup as a well-documented test-case to demonstrate how the proposed deep LSMC methodology scales to multiple sources of uncertainty.}
We exemplify this approach with an application to the DICE model with uncertainties in: (1) equilibrium climate sensitivity, (2) the damage function coefficient, (3) the growth rate of total factor productivity, (4) the growth rate of decarbonization, and (5) the equilibrium carbon concentration in the upper strata, {\color{black}where (3)–(4) enter as stochastic process shocks and (1)–(2), (5) are treated as time-invariant (initial) parameter uncertainty.}
These five uncertainties were identified in \citet{nordhaus2018} as being major sources of uncertainty for the evolution of climate-economic state variables.
Typically, polynomial regression is used in LSMC to approximate the corresponding conditional expectations with respect to state variables and controls.
However, for models such as the stochastic DICE model, this leads to the need for too many covariates and simulations, making the method not practical.
To overcome this problem, we use deep neural network approximations for the required regressions and provide detailed explanations.
{\color{black}In addition to computing optimal feedback policies, we use the deep LSMC solution to perform uncertainty quantification for key quantities of interest (e.g., the social cost of carbon and temperature
trajectories) and variance-based global sensitivity analysis via Sobol' indices; see Section~\ref{section-numerics}.}

The DICE model is a deterministic approach that combines a Ramsey--Cass--Koopmans neoclassical model of economic growth (also known as the Ramsey growth model) with a simple climate model.
It involves six state variables (economic capital; temperature in atmosphere and lower oceans; carbon concentration in atmosphere, upper and lower oceans) evolving deterministically in time, two control variables (savings and carbon emission reduction rates) to be determined for each time period of the model, and several exogenous processes (e.g. population size and technology level).
The uncertainty about the future of the climate and economy is then typically assessed by treating some model parameters as random variables (because we do not know the exact true value of the key parameters) using a Monte Carlo analysis (see \cite{nordhaus2018,gillingham2015modeling}).

Modeling uncertainty owing to the stochastic nature of the state variables (i.e.\ owing to the process uncertainty that is present even if we know the model parameters exactly) requires the development and solution of the DICE model as a dynamic model of decision-making under uncertainty, where we calculate the optimal policy response under the assumption of continuing uncertainty throughout the time frame of the model.
Few attempts have been made to extend the DICE model to incorporate stochasticity in the underlying state variables and solve it as a recursive dynamic programming problem.
For example, \citet{kelly1999} and \citet{leach2007} formulated the DICE model with stochasticity in the temporal evolution of temperature, and solved this as a recursive dynamic programming problem.
These studies are seminal contributions to the incorporation of uncertainty in the DICE model (although their numerical solution approach is difficult to extend to a higher dimensional space and time-frequency).
\cite{cai2019} formulate DICE as a dynamic programming problem with a stochastic shock on the economy and climate.
In addition, \cite{traeger2014} developed a reduced DICE model with a smaller number of state variables, whereas \cite{lontzek2015stochastic} studied the impact of climate tipping points, and \cite{shevchenko2022impact} considered the DICE model with discrete stochastic shocks to the economy.
To the best of our knowledge, the only attempt to solve the stochastic DICE model using an LSMC-type approach is \cite{ikefuji2020expected}.
Their study handles only one uncertainty at a time, and the setup of the regression type Monte Carlo algorithm omits the integration for the conditional expectation in the Bellman equation, assuming the randomness is known in the transition of state variables (in principle, in this case, the required integration can be performed by using deterministic quadrature methods, but this will be subject to the curse of dimensionality).

The primary contributions of our paper are as follows:
\begin{enumerate}
\item We introduce an efficient approach for modeling stochastic climate-economy models by combining the least-squares Monte Carlo method with deep learning techniques. It provides flexibility in handling various types of uncertainties, including both parametric and stochastic process uncertainties.
\item We formulate a stochastic version of the DICE model using the sources of uncertainty as identified by \citet{nordhaus2018}. Notably, it does not rely on discretizing the underlying probability distributions, as is usually done in Monte-Carlo-type analyses for the sake of model tractability.
\item {\color{black}We perform comprehensive numerical experiments and discuss uncertainty quantification (UQ) for key quantities of interest. Our variance-based global sensitivity analysis (GSA) using Sobol' indices is closely related to modern UQ/GSA approaches in economics and computational economics \citep{harenberg2019uqgsa,scheidegger2019mlhd} and, in particular, to the DICE sensitivity analysis of \citet{miftakhova2021gsa}.}
\end{enumerate}

{\color{black} %
The main contribution of this paper is methodological: we propose a deep least-squares Monte Carlo solver for recursive climate-economy stochastic control problems with endogenous state dynamics, inequality-constrained policies, and multiple uncertainty sources.\footnote{The code used for the numerical experiments is available upon request.} We use the DICE-2016R2 calibration and the five uncertainty blocks reported in \citet{nordhaus2018} as a transparent benchmark to facilitate comparison with the IAM literature, not because we view this particular uncertainty specification as uniquely policy-relevant or definitive. The deep LSMC algorithm only requires a simulatable one-step transition map; therefore, alternative uncertainty blocks, recalibrations, or climate-emulator variants can be incorporated without conceptual changes to the backward-recursion framework. For practical guidance on when deep LSMC is preferable to grid / VFI / projection approaches (and when it is not), see Section~\ref{subsection-comparison}.

While the present paper is primarily methodological, the stochastic-DICE output distributions produced by our framework (e.g.\ temperature paths, the SCC, and policy trajectories under uncertainty) can be used as inputs to climate-finance and actuarial risk assessments. For instance, \citet{arandjelovicShevchenko2025netzero} quantify uncertainty in the SCC and related tail-risk measures under net-zero scenarios, and \citet{arandjelovicShevchenko2025reserves} develop a stress-testing framework for life-insurance reserving driven by stochastic-DICE temperature projections.}

The paper is organized as follows.
Section~\ref{section-model} gives a description of the considered model.
Section~\ref{section-LSMC} describes the numerical method used to solve the model.
Section~\ref{section-numerics} provides a comprehensive numerical study.
Section~\ref{section-conclusion} concludes.

\section{Model description}\label{section-model}
In this section, we present the DICE-2016R2 model as a classical example of a recursive climate-economy model.
This version of the DICE model was used in \citet{nordhaus2018}.
It includes parameter uncertainties in equilibrium climate sensitivity, the damage function coefficient and the {\color{black} \emph{carbon-cycle coefficient} (equilibrium carbon concentration in the biosphere/upper ocean strata)}, as well as process uncertainties in the growth rate of total factor productivity and the growth rate of decarbonization.

The original deterministic DICE model seeks to find policies $\pi$ that maximize a social welfare function, which models the discounted sum of population-weighted utility of per capita consumption:
\begin{equation}\label{eq:baseDICE}
V_0 = \sup_{\pi} \sum_{t=0}^{\infty} \rho^{t} L_{t}\, u\left(\frac{C_t}{L_t}\right),
\end{equation}
where $\rho\in(0,1)$ is the $\Delta$-step discount factor (in the DICE-2016R2 calibration the annual pure rate of time preference is $0.015$, hence $\rho=(1+0.015)^{-\Delta}$), $L_{t}$ is world population, $C_t$ is total consumption, and the time index $t=0,1, \hdots$ corresponds to $\Delta=5$-year steps.
{\color{black}In the numerical implementation (as in standard DICE codes), we approximate this by a long but finite horizon with a terminal reward; see the discussion around Table~\ref{model-parameters} and Section~\ref{section-numerics}.}
We parameterize $\pi_t=(c_t,\mu_t)$, where $c_t\in[0,1]$ is the \emph{consumption share} of net output $Q_t$ (so that $C_t=c_t Q_t$ and per-capita consumption equals $C_t/L_t$) and $\mu_t\in[0,1]$ is the industrial-emissions mitigation rate.
The utility function $u$ has constant relative risk aversion, $u(c)=\frac{c^{1-\alpha}-1}{1-\alpha}$, with risk-aversion parameter $\alpha\ge 0$ (the case $\alpha=1$ corresponds to logarithmic utility).

{\color{black}%
\begin{remark}\label{rem:risk_sensitive_prefs}
The baseline DICE specification uses CRRA utility. While convenient and standard in the DICE benchmark, CRRA ties together the coefficient of relative risk aversion and the elasticity of intertemporal substitution (IES), which can attenuate precautionary motives and thereby mute the observable impact of uncertainty on optimal policies and the social cost of carbon (SCC). To separate risk aversion and intertemporal substitution, one may instead adopt recursive preferences such as Epstein--Zin utility \citep{epstein1989substitution,weil1990nonexpected}, as used in stochastic IAM analyses including \citet{cai2019}. Our deep LSMC approach extends directly to this case because the Bellman recursion remains backward and regression-based: the continuation step requires approximating conditional expectations of a (power-transformed) continuation value.
\end{remark}
}

The deterministic DICE core features six endogenous state variables: economic capital $K_{t}$, the concentration of carbon in the atmosphere, the upper oceans, and the lower oceans, $M_{t} = (M_{t}^{\mathrm{AT}}, M_{t}^{\mathrm{UP}}, M_{t}^{\mathrm{LO}})^{\top}$, and the global mean temperature of the Earth's surface and the deep oceans, $T_{t} = (T_{t}^{\mathrm{AT}}, T_{t}^{\mathrm{LO}})^{\top}$.
In our stochastic specification (Subsection~\ref{subsection-uncertainty}), the Markov state used in the numerical recursion is augmented by the realized exogenous levels $(A_t,\sigma_t)$ and by the three uncertain structural parameters (drawn at $t=0$ and held fixed along a trajectory); see Section~\ref{section-numerics} for the resulting 11-dimensional state vector used in the numerical study.
The evolution of the economic and geophysical sectors is governed by the dynamics described below.

\textbf{The economic system:} Gross output is modeled by a Cobb--Douglas production function of capital, labor, and technology, $Y_{t} = A_{t} K_{t}^{\gamma}L_{t}^{1-\gamma}$, where $\gamma \in (0,1)$ and $1-\gamma$ are the output elasticities of capital and labor, respectively.
Here, $A_{t}$ denotes \emph{total factor productivity} (see Subsection \ref{subsection-uncertainty}), representing technological progress and efficiency improvements over time.

The DICE model incorporates economic damages from climate change, represented by a damage function that is quadratic in the global mean surface temperature, $d_{t} = \pi_{2}\times (T_{t}^{\mathrm{AT}})^{2}$, where $\pi_{2}$ is the \emph{damage coefficient} (see Subsection \ref{subsection-uncertainty}).
These damages can be mitigated by emission reduction, controlled by the policy $\mu_{t}$.
Reducing emissions incurs abatement costs $\Lambda_{t}$ (see Table~\ref{model-parameters} for their specification).

Net output is then given by gross output reduced by damages and abatement costs, $Q_{t} = (1-\Lambda_{t})Y_{t} / (1+d_{t})$, and economic capital $K_{t}$ evolves according to the following dynamics:
\begin{equation}
K_{t+1} = (1-\delta_{K})^{\Delta}K_{t} + \Delta \times (Q_{t} - C_{t}),
\end{equation}
where $\delta_{K}$ is the rate of depreciation of economic capital.
In terms of the consumption-share control $c_t$ defined above, total consumption is $C_t=c_t Q_t$, so that investment equals $(1-c_t)Q_t$ and the capital update can be written as $K_{t+1} = (1-\delta_{K})^{\Delta}K_{t} + \Delta \times (1-c_t)Q_t$.

\textbf{The carbon cycle:} The carbon cycle is modeled by three reservoirs, which follow the dynamics:
\begin{equation}
M_{t+1} = \Phi M_{t} +  (\Delta\times\beta E_{t}, 0, 0)^{\top},
\end{equation}
where $\Phi$ is a coefficient matrix, $E_{t}$ is total $\mathrm{CO}_{2}$ emissions (in billions of tons per year), and $\beta$ is conversion factor of $\mathrm{CO}_{2}$ mass into the equivalent mass of carbon.
Emissions $E_{t}$ are equal to uncontrolled industrial emissions, given by a level of \emph{carbon intensity} (see Subsection \ref{subsection-uncertainty}) $\sigma_{t}$ times gross output, reduced by the emission reduction rate $\mu_{t}$, plus exogenous land-use emissions $\widetilde{E}_{t}$, i.e.\ $E_{t} = \sigma_{t} (1-\mu_{t}) Y_{t} + \widetilde{E}_{t}$.

\textbf{The temperature module:} The relationship between greenhouse gas accumulation and increased radiative forcing is described by the function:
\begin{equation*}
F_{t} = \eta \log_{2}\big( M_{t}^{\mathrm{AT}} / \widetilde{M}^{\mathrm{AT}} \big) + \widetilde{F}_{t},
\end{equation*}
which models the change in total radiative forcings from anthropogenic sources such as $\mathrm{CO}_{2}$.
It consists of exogenous forcings $\widetilde{F}_{t}$ plus forcings due to atmospheric concentrations of $\mathrm{CO}_{2}$.
Here, $\widetilde{M}^{\mathrm{AT}}$ is the preindustrial atmospheric carbon concentration.
The evolution of global mean temperatures follows the dynamics:
\begin{equation}
T_{t+1} = \Psi T_{t} + \bigg(\begin{array}{c} \psi_{1}F_{t+1} \\ 0 \end{array}\bigg),
\end{equation}
where $\Psi$ is a coefficient matrix, and $\psi_{1}$ is a model parameter.
It is important to note that $T_{t}$ is measured in terms of the absolute increase in temperature relative to the year 1900.

{\color{black}
\begin{remark}\label{rem:dice-climate-calibration}
Recent work by \citet{dietz2021climatedynamics} and \citet{folini2025climate} shows that the \emph{benchmark} climate block in DICE-2016/DICE-2016R2 does not fully reproduce the behavior of more detailed Earth-system models for key physical benchmarks, including aspects of the transient temperature response and related ocean heat-uptake dynamics. It was demonstrated in \cite{folini2025climate} that the functional form
of the climate variable equations in DICE-2016 is fit for purpose and can be re-calibrated to match the benchmark data. In this paper we retain the DICE-2016R2 calibration as a widely used benchmark in order to facilitate comparison with the existing IAM literature and to focus on the numerical contribution. Accordingly, quantitative results should be interpreted conditional on this benchmark climate block. At the same time, the deep LSMC method developed below is not tied to this particular emulator: it requires only a simulatable one-step transition map and can therefore be combined with alternative (climate-science-consistent) climate emulators, including re-calibrated DICE variants as in \citet{folini2025climate}.
\end{remark}
}

In DICE-2016R2, $\mu_{t}$ is assumed to be non-negative with an upper bound of 1, i.e.\ no negative industrial emissions are allowed.
Table \ref{model-parameters} summarizes the main coefficients of the model.
Note that the number of time steps $N$ is chosen such that $t=0$ corresponds to the year 2015, while $t=N$ corresponds to the year 2500.

{\color{black}%
\begin{remark}\label{rem:ct-link}
For readers more familiar with continuous-time stochastic control, the discrete-time DICE transition equations can be viewed as arising from sampling (and discretising) an underlying continuous-time controlled system on calendar time $s$ and then evaluating it at a grid with step size $\Delta=5$ years.

For instance, a continuous-time capital-accumulation dynamics of the form
\begin{equation*}
\mathrm{d}K_s = \big(Q_s - C_s - \delta_K K_s\big)\,\mathrm{d}s \;+\; \nu_K(X_s,\pi_s)\,\mathrm{d}W_s,
\end{equation*}
(with state vector $X_s$, Brownian motion $W_s$, and $\nu_K\equiv 0$ in the deterministic DICE-2016R2 calibration) yields, after holding controls approximately constant over each $\Delta$-year period and discretising in time, a one-step update of the form used in the discrete-time model. The carbon-cycle and temperature blocks admit analogous continuous-time representations, with discrete-time innovations interpretable as aggregated shocks over a $\Delta$-year period.

We retain the standard $\Delta$-step DICE-2016R2 specification in order to preserve the benchmark calibration and facilitate comparison with the IAM literature. The deep LSMC method developed in Section~\ref{section-LSMC} only requires that the resulting one-step transition map can be simulated, and therefore applies equally to time-discretised continuous-time climate-economy models; see also \citet{cai2012} for a continuous-time integrated-assessment formulation.
\end{remark}
}

\textbf{The social cost of carbon (SCC):} The social cost of carbon (SCC) is a measure of the economic harm caused by emitting one additional ton of carbon dioxide ($\mathrm{CO}_{2}$) into the atmosphere.
It represents the present value of the damages associated with a marginal increase in $\mathrm{CO}_{2}$ emissions in a given year.
The SCC is typically expressed in monetary terms (e.g. dollars per ton of $\mathrm{CO}_{2}$) and is used to help policymakers evaluate the benefits of reducing emissions and compare the costs of different climate policies or regulatory actions aimed at mitigating climate change.
The SCC can be calculated in the DICE model by:
\begin{equation}\label{equation-SCC}
SCC_{t} = -1000 \beta \frac{\partial V_{t} /\partial M_{t}^{\mathrm{AT}}} {\partial V_{t} / \partial K_{t}},
\end{equation}
where $V_{t}$ denotes the value function at time $t$, and $\beta$ represents the $\mathrm{CO}_{2}$ to carbon mass transformation coefficient.
Here $K_t$ is measured in trillions of 2010 USD and $M_{t}^{\mathrm{AT}}$ in gigatons of carbon, so $(\partial V_{t}/\partial M_{t}^{\mathrm{AT}})/(\partial V_{t}/\partial K_{t})$ has units
``trillion USD per GtC''; multiplying by $1000$ converts to USD per tC, and $\beta=1/3.666$ converts from tC to tCO$_2$.

\begin{table}[htb!]
\centering
\begin{threeparttable}
\caption{Parameters for the base model.\label{model-parameters}}
\begin{tabular}{l}
\toprule
$N=97$ time steps of $\Delta=5$ years\\
$L_{t+1} = L_{t}(11.500 / L_{t})^{0.134}$, $L_{0}=7.403$ (in billions)\\
$A_{t+1} = A_{t} / (1-g_{A}(t))$, $g_{A}(t+1) = g_{A}(t) \exp(-0.005 \Delta)$, $A(0) = 5.115$, $g_{A}(0) = 0.076$\\
$\sigma_{t+1} = \sigma_{t} \exp(g_{\sigma}(t)\Delta)$, $g_{\sigma}(t+1) =g_{\sigma}(t)(1-0.001)^{\Delta}$, $\sigma_{0} = \frac{35.85}{105.5(1-0.03)}$, $g_{\sigma}(0) = -0.0152$\\
$\Lambda_{t} = 550 (1-0.025)^{t} \sigma_{t} \mu_{t}^{\theta_{2}}/ (1000\theta_{2})$\\
$\widetilde{E}_{t}=2.6(1-0.115)^{t}$, $\widetilde{F}_{t} = (0.5 + t/34)\indicatorset{t < 17} + \indicatorset{t \ge 17}$\\
$K_{0}= 223$, $M_{0}^{\mathrm{AT}}=851$, $M_{0}^{\mathrm{UP}}=460$, $M_{0}^{\mathrm{LO}}=1740$, $T_{0}^{\mathrm{AT}}=0.85$, $T_{0}^{\mathrm{LO}}=0.0068$\\
$\alpha=1.45$, $\beta=1/3.666$, $\gamma=0.3$, $\rho=(1+0.015)^{-\Delta}$, $\delta_{K} = 0.1$\\
$\Phi=\left(\begin{array}{ccc} \phi_{11} & \phi_{12} & 0 \\ \phi_{21} & \phi_{22} & \phi_{23} \\ 0  &  \phi_{32} & \phi_{33}\\ \end{array}\right)$,
$\Psi=\left(\begin{array}{cc} 1-\psi_1\psi_2-\psi_1\psi_3 & \psi_1\psi_3 \\ \psi_4 & 1-\psi_4 \\ \end{array} \right)$\\
$\phi_{21}=0.12$, $\phi_{32}=0.007$, $\phi_{11} = 1-\phi_{21}$, $\phi_{12}=\phi_{21}588/360$\\
$\phi_{22}=1-\phi_{12}-\phi_{32}$, $\phi_{23}=\phi_{32}360/1720$, $\phi_{33}=1-\phi_{23}$\\
$\psi_{4}=0.025$, $\psi_{1}=0.1005$, $\psi_{3}=0.088$, $\psi_{2}=3.6813/3.1$\\
$\eta=3.6813$, $\widetilde{M}^{\mathrm{AT}}=588$, $\pi_{2}=0.00236$, $\theta_{2}=2.6$\\
\bottomrule
\end{tabular}
\end{threeparttable}
\end{table}

\subsection{Modeling uncertainty}\label{subsection-uncertainty}
The dynamics presented in the DICE model so far are purely deterministic, assuming precise knowledge of the future evolution of all exogenous variables for centuries ahead.
This approach is an unrealistic simplification.
A reasonable way to address this issue is to introduce probabilistic distributions into the model to account for uncertainties about future outcomes.
In this paper, we distinguish between two types of uncertainties: stochastic process uncertainty, and initial parameter uncertainty.

\textit{Stochastic process uncertainty} refers to the uncertainty in the evolution of future trajectories of exogenous variables.
A classical example from quantitative finance is Brownian motion, $B = (B_{t})_{t \ge 0}$, modeled by $B_{0} = 0$ and $B_{t+h} - B_{t} \sim \mathcal{N}(0, h)$ for $t \ge 0$ and $h \ge 0$, where $\mathcal{N}(0, h)$ denotes the normal distribution with expected value $0$ and variance $h$.
Incorporating stochastic process uncertainties is challenging because the uncertainty propagates over time, increasing the volatility of the variable's distribution.
The LSMC method we present below is highly sensitive to introduced volatility, making this incorporation a significant challenge that few contributions in the climate-economy literature have successfully addressed.

\textit{Initial parameter uncertainty} refers to uncertainty about one or more parameters in the system that remain fixed over time.
A common method to study this uncertainty is a perturbation analysis, where parameters are sampled, the model is solved, and the process is repeated.
However, this approach does not accurately depict the model's evolution over time, as an agent in the model would consider overall outcome uncertainty, not individual instances of the uncertain parameter.
Another related concept is Bayesian learning \citep{kelly1999}, where the parameter distribution evolves over time as more information about the system is revealed.
This type of uncertainty can be treated by the LSMC approach presented in this paper, but we chose not to include this in the current study, leaving it for future work.

Identifying reasonable uncertainties to include in the model is challenging, as some uncertainties might be more significant than others.
Advanced statistical analyses are required to make educated assumptions about probability distributions for the climate and economic system.
For our paper, we incorporate five uncertainties into the DICE model, as identified by \citet{nordhaus2018}.
These include stochastic process uncertainties in the growth rates of total factor productivity $A$ and the rate of decarbonization $\sigma$, as well as initial parameter uncertainties in the temperature-sensitivity coefficient, the damage coefficient, and the carbon cycle coefficient.
{\color{black}Other uncertainty dimensions studied in the IAM literature include structural/model uncertainty and ambiguity / robust-control formulations; see, e.g., \citet{barnett2020pricing}.}
We emphasize that our method is not limited to the chosen uncertainties, and we now explain our choices in more detail.
Unless stated otherwise, we assume that the two process shocks driving productivity and decarbonization are independent across time and mutually independent, and that they are independent of the three time-$0$ parameter draws. The parameter draws are realized at $t=0$ and then held fixed along each simulated trajectory.

{\color{black}%
\begin{remark}
We adopt the parametric uncertainty inputs proposed by \citet{nordhaus2018} primarily as a transparent benchmark that facilitates comparison with the existing stochastic-DICE literature, and not as a claim that these inputs represent a definitive or up-to-date probabilistic calibration. Recent IPCC assessments summarize physical-science evidence largely in terms of assessed ranges and scenario ensembles (SSP/RCP), rather than as a ready-to-use probabilistic calibration of reduced-form IAM parameters. Mapping IPCC assessments to a stochastic DICE specification typically requires additional modeling choices (e.g., selecting a parametric family, matching assessed quantiles, and maintaining internal consistency of the reduced-form climate emulator when varying physical parameters). In this paper we therefore (i) retain the benchmark specification for our main numerical experiments, and (ii) explicitly compare key benchmark inputs to AR6 assessed ranges where a direct comparison is meaningful (see the discussion of ECS below and \citet{ipcc2021wgi_spm}).

Finally, SSP/RCP scenario analysis is naturally viewed as a complementary ``between-scenario'' layer: one may recalibrate the exogenous trajectories (population, productivity, emissions intensity) to a given SSP pathway and then re-solve the resulting stochastic control problem with the same deep LSMC algorithm; see, e.g., \citet{riahi2017ssp} and the ScenarioMIP/CMIP6 overview \citet{oneill2016scenariomip}.
\end{remark}
}

\begin{figure}[htb!]
\centering
\captionsetup{width=0.75\textwidth,format=plain}
\includegraphics[width=0.75\linewidth]{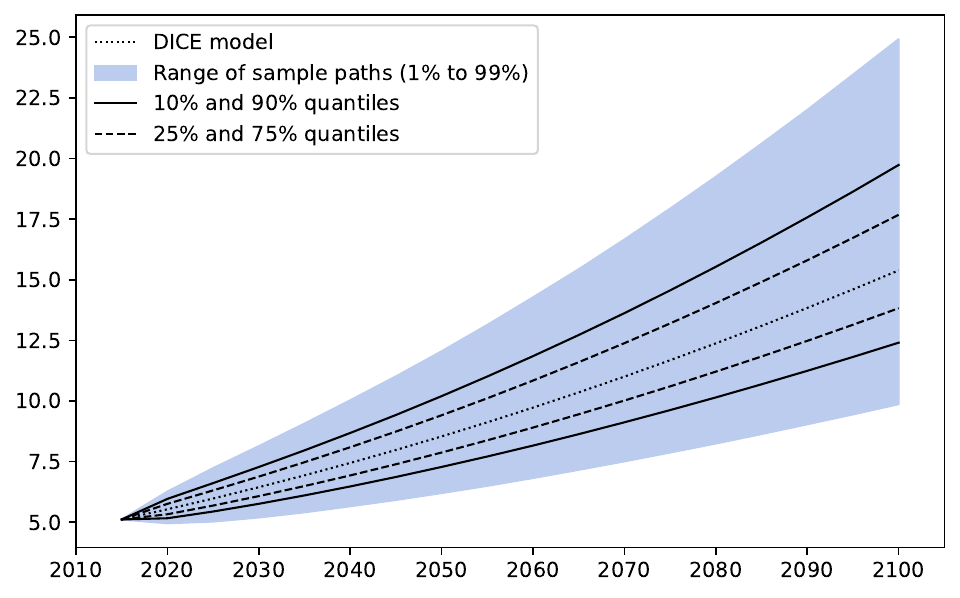}
\vspace*{-0.40cm}
\caption{Evolution of total factor productivity $A$ under the assumption that the growth rate $g_{A}$ is uncertain.}
\label{tfp-evolution}
\end{figure}

\textbf{Productivity growth.} Assuming a Cobb-Douglas production function, the growth in total factor productivity $A$ models the growth in output that is not explained by growth in inputs of labor and capital used in production.
The DICE model assumes $A$ evolves according to $A_{t+1} = A_{t}/(1-g_{A}(t))$, where $g_{A}(t)$ is the deterministic growth rate which is specified in Table~\ref{model-parameters}.
\citet{nordhaus2018} assumes $g_{A}(0)$ is normally distributed with mean $0.076$ and standard deviation $0.056$.
But in this case, using the dynamics for the growth rate, we can model $g_{A}(t)$ as normally distributed with mean $g_{A}(0) \exp(-0.005 t \Delta)$ and standard deviation $0.056 \exp(-0.005 t \Delta)$.
In order to remove extreme cases, we truncate this distribution at the mean $\pm$ two standard deviations.
The evolution of $A_{t}$ is shown in Figure~\ref{tfp-evolution}.

\begin{figure}[htb!]
\centering
\captionsetup{width=0.75\textwidth,format=plain}
\includegraphics[width=0.75\linewidth]{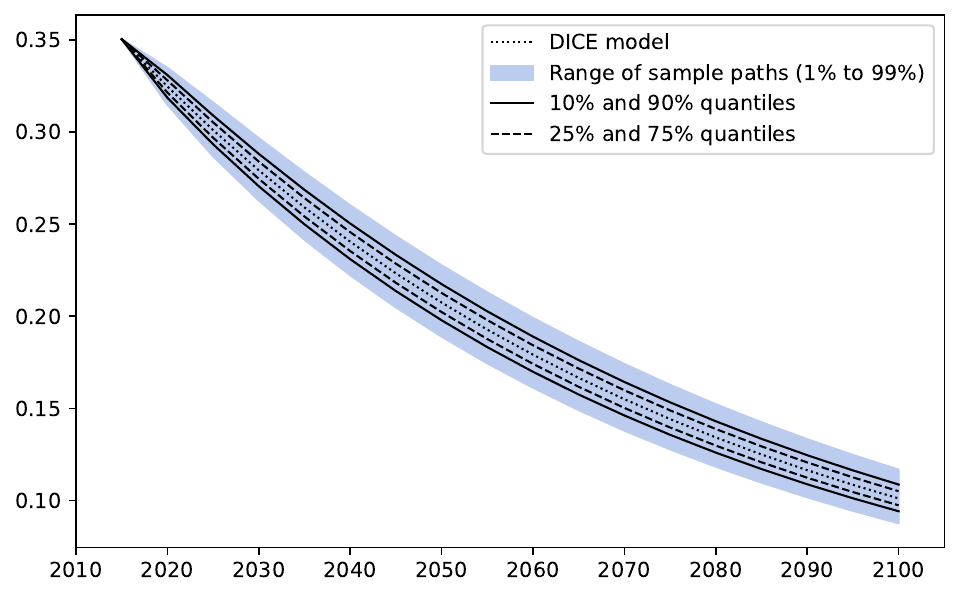}
\vspace*{-0.40cm}
\caption{Evolution of carbon intensity $\sigma$ under the assumption that the growth rate $g_{\sigma}$ is uncertain.}
\label{sigma-evolution}
\end{figure}

\textbf{The rate of decarbonization.} Uncontrolled industrial $\mathrm{CO}_{2}$ emissions are given by a level of carbon intensity, $\sigma_{t}$, times gross output.
The DICE model assumes $\sigma$ evolves according to $\sigma_{t+1} = \sigma_{t} \exp(g_{\sigma}(t) \Delta)$, with a deterministic growth rate $g_{\sigma}(t)$ which is specified in Table~\ref{model-parameters}.
\citet{nordhaus2018} assumes $g_{\sigma}(0)$ is normally distributed with mean $-0.0152$ and standard deviation $0.0032$.
We therefore model $g_{\sigma}(t)$ as normally distributed with mean $g_{\sigma}(0) (1-0.001)^{t \Delta}$ and standard deviation $0.0032 (1-0.001)^{t \Delta}$, truncating the distribution at the mean $\pm$ two standard deviations in order to remove extreme cases.
The evolution of $\sigma_{t}$ is shown in Figure~\ref{sigma-evolution}.

{\color{black}
\textbf{Equilibrium climate sensitivity (ECS).}
Equilibrium climate sensitivity (ECS) measures the long-run increase in global mean surface temperature following a doubling of atmospheric $\mathrm{CO_2}$.
The DICE-2016R2 benchmark corresponds to ECS $= 3.1 \degree \mathrm{C}$ for an equilibrium $\mathrm{CO_2}$ doubling (see Table~\ref{model-parameters} and the denominator in the definition of $\psi_{2}$).
\citet{nordhaus2018} models ECS as a log-normal distribution, $\exp(X)$ with $X \sim \mathcal{N}(1.1060, 0.2646^{2})$, and we follow this benchmark specification, truncating at the mean $\pm$ two standard deviations.

To relate this benchmark to the current physical-science assessment, the IPCC AR6 Working Group~I Summary for Policymakers reports an assessed best estimate of ECS of $3 \degree\mathrm{C}$, with a likely range of $2.5$--$4 \degree\mathrm{C}$ and a very likely range of $2$--$5 \degree\mathrm{C}$ \citep[SPM A.4.4]{ipcc2021wgi_spm}.
Under the benchmark distribution above (with the stated truncation), the implied median is close to $3 \degree\mathrm{C}$ and the upper tail is somewhat heavier than the AR6 ``likely'' range.
We stress that ECS is an exogenous calibration input in reduced-form IAMs such as DICE; alternative (e.g., IPCC-informed) probabilistic specifications can be substituted without changing the deep LSMC methodology.

Because ECS enters through the reduced-form DICE climate emulator, varying ECS should ideally be accompanied by a consistent re-calibration of the remaining climate-module parameters (e.g., feedback and heat-uptake coefficients); see \citet{folini2025climate} for discussion. In our benchmark numerical study we follow \citet{nordhaus2018} by varying ECS through the DICE climate-sensitivity parametrization while keeping the remaining emulator coefficients at their DICE-2016R2 values; our UQ results should be interpreted conditional on this benchmark calibration.}

\textbf{The damage function.} The DICE model assumes that climate-induced economic damages are a quadratic function of the increase in atmospheric temperature.
It is modeled as a fractional loss of global output from greenhouse warming, $d_t = \pi_{2} \times(T_{t}^{\mathrm{AT}})^{2}$, where $\pi_{2}$ denotes a damage coefficient representing the severity of the economic impact of global warming.
The DICE model assumes $\pi_{2}$ to be equal to 0.00236.
\citet{nordhaus2018} models the $\pi_{2}$ by a normal distribution with mean 0.00236 and standard deviation 0.00118.
We use the same distribution but truncate it at the mean minus one standard deviation, and at the mean plus two standard deviations.

{\color{black}The functional form and calibration of damages in IAMs are actively debated, and alternative specifications can imply materially different SCC levels and policy prescriptions; see, e.g.,
\citep{botzen2012how,carleton2022valuing,dietz2024iam}. Our contribution here is methodological: the deep LSMC recursion applies unchanged under alternative damage specifications and recalibrations.}

\textbf{The carbon cycle.} The carbon cycle coefficient models the equilibrium concentration of carbon in the biosphere and upper level of the oceans.
The DICE model assumes it to be equal to 360 gigatonnes of carbon (GtC).
In Table~\ref{model-parameters}, it corresponds to the value 360 appearing in the definitions of $\phi_{12}$ and $\phi_{23}$.
\citet{nordhaus2018} models this coefficient as a log-normal distribution, $\exp(X)$ with $X \sim \mathcal{N}(5.8510, 0.2649^{2})$.
We do the same, truncating at the mean $\pm$ two standard deviations.

\begin{figure}[htb!]
\centering
\captionsetup{width=0.85\textwidth,format=plain}
\includegraphics[width=0.85\linewidth]{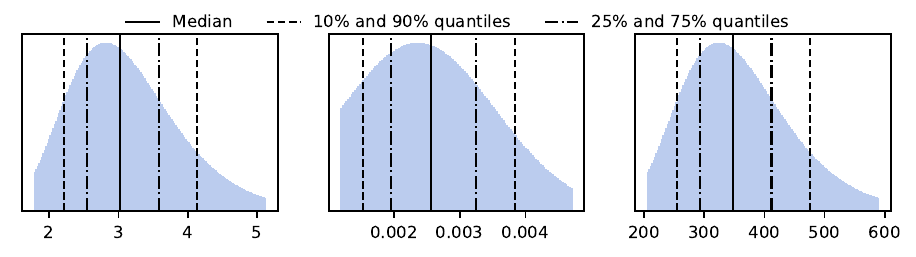}
\vspace*{-0.40cm}
\caption{Density plots of the parameter distributions of equilibrium climate sensitivity (left panel), the damage coefficient (middle panel), and carbon cycle coefficient (right panel).}
\end{figure}

\begin{remark}\strut
\begin{itemize}
    \item Another type of uncertainty is \textit{parametric uncertainty}, where the value of a coefficient can change over time as it is re-drawn at each point in time.
This type of uncertainty lies between the stochastic process and the initial parameter uncertainty.
Although we did not include it in our study, it is straightforward to incorporate and solve using our method.
\item
Assuming $\pi_{2} \sim \mathcal{N}(0.00236, 0.00118^{2})$ implies a roughly $2.3 \%$ probability of $\pi_{2}$ being negative.
This is a non-negligible scenario.
Given that the DICE model aims to combine equations for the economy and climate, it is highly questionable to assume the damage coefficient could be below or just above zero.
Moreover, the assumption of a log-normal distribution for the equilibrium climate sensitivity and the carbon cycle coefficient also entails a non-negligible probability of those coefficients being close to zero.

\citet{nordhaus2018} avoids this issue by discretizing the distributions, separating them into quintiles, and then calculating the expected values of the random variables within those quintiles.
These expected values are taken as realizations of discrete uncertain variables, yielding sufficiently positive lowest realizations for the coefficients.
Inspired by this approach, we also truncate the distributions of the random variables, however, without discretizing them.
All quantitative results below should be interpreted conditional on this benchmark truncation, which avoids issues with too low damage coefficients and temperature sensitivities, as well as extreme growth rates for total factor productivity and carbon intensity.
\end{itemize}
\end{remark}

\section{The deep least-squares Monte Carlo method}\label{section-LSMC}
The numerical solution of the model is achieved using the endogenous state least-squares Monte Carlo (LSMC) algorithm with control randomization, as introduced by \cite{kharroubi2014numerical} and adapted for expected utility optimal stochastic control problems by \cite{andreasson2022}.
This method approximates the conditional expectation of the value function in the Bellman equation using regression with a quadratic loss function applied to the transformed value function.
Typically, regression basis functions are ordinary polynomials of the state and control variables, usually up to the third order.
In our implementation, we use \emph{deep neural networks} to approximate the regression predictor.
To mitigate transformation bias in the regression estimate of the conditional expectation, we employ the smearing estimate (see \cite{Duan1983Smearing,andreasson2022}).

{\color{black} %
\begin{remark}
The transformation-bias adjustment and smearing-based correction used in our regression step (see Subsection~\ref{subsection-biastransform}) follow the approach of \citet{andreasson2022}. Our methodological contribution here is the extension of this baseline to a stochastic climate-economy control problem with continuous, inequality-constrained policies and multiple simultaneous uncertainty sources: we use deep neural networks as the regression class, and we incorporate IAM-specific design choices (post-decision-state formulation, normalization, and low-discrepancy sampling) that materially improve numerical stability and precision in the stochastic DICE setting.
\end{remark}
}

Let $t=0, 1, \hdots, N$ correspond to time points in the interval $[0, T]$.
Consider the standard discrete dynamic programming problem with the objective to maximize the expected value of the utility-based total reward function
\begin{equation}\label{eq:framework_valuefunction}
V_{0}(x) = \sup_{\pi} \mathbb{E} \bigg[ \sum_{t=0}^{N-1} \rho^{t} R_{t}(X_{t}, \pi_{t}) + \rho^{N} R_{N}(X_{N})\, \bigg|\, X_{0} = x; \pi \bigg],
\end{equation}
where $\pi = (\pi_{t})_{t=0, 1, \hdots, N-1}$ is a control, $X = (X_{t})_{t=0, 1, \hdots, N}$ is a controlled state variable, $R_{N}$ and $R_{t}$ are reward functions, $\rho$ is a time discount factor, and the expectation is conditional on the initial state $X_{0}=x$ and following the policy $\pi$.
The evolution of the state variable is specified by a transition function $\mathcal{T}_t(\cdot)$ such that
\begin{equation}
\label{eq:transitionfunction}
X_{t+1} = \mathcal{T}_t(X_t, \pi_t, Z_{t}),
\end{equation}
where $Z_{0}, Z_{1}, \hdots, Z_{N-1}$ are independent disturbance terms, i.e.\ the state of the next period depends on the current state's value,  the current period's control decision, and the realization of the disturbance term.
{\color{black}%
This discrete-time representation also covers models specified originally in continuous time: after a time discretisation of a controlled SDE one obtains a one-step transition of the form \eqref{eq:transitionfunction}, with $Z_t$ representing (for example) Brownian increments over a time step.%
}

This problem can be solved using the backward recursion of the Bellman equation, starting from $V_{N}(x) = R_{N}(x)$ and then solving recursively:
\begin{equation}\label{eq:bellmanEquation}
V_{t}(x) = \sup_{\pi_{t} \in \mathcal{A}_{t}} \bigg\{ R_{t}(x, \pi_{t}) + \mathbb{E}\big[\rho V_{t+1}(X_{t+1})\, \big|\, X_{t} = x; \pi_{t} \big] \bigg\}, \quad t=N-1, N-2, \hdots, 0,
\end{equation}
where the expectation is conditional on the state $X_{t}=x$ and the policy $\pi_{t}$ at time $t$.
For further details on dynamic programming, we refer the interested reader to the excellent monograph by \citet{fleming06} on the subject.

Using Equation \eqref{eq:bellmanEquation}, the optimal control can be found by solving:
\begin{equation}\label{optimal-policy}
\pi_{t}^{\ast}(x) = \argsup_{\pi_{t} \in \mathcal{A}_t} \bigg\{ R_{t}(x, \pi_{t}) + \mathbb{E}\big[\rho V_{t+1}(X_{t+1})\, \big|\, X_{t} = x; \pi_{t} \big] \bigg\}.
\end{equation}
Here, $\mathcal{A}_t$ denotes a set of admissible values of $\pi_{t}$, which may depend on $x$.
When the number of stochastic variables is more than three, it usually becomes computationally infeasible to use quadrature-based methods to evaluate the conditional expectation in \eqref{eq:bellmanEquation}, making simulation methods like LSMC preferable.
The increase in the number of state variables will call for LSMC too.

The LSMC method approximates the conditional expectation in equation \eqref{eq:bellmanEquation}:
\begin{equation}\label{eq:conditionalexp}
\Phi_t(X_{t},\pi_t) = \mathbb{E}\big[\rho V_{t+1}(X_{t+1})\, \big|\, X_{t}; \pi_{t} \big]
\end{equation}
using a regression scheme with the states $X_{t}$ and randomized policies $\pi_{t}$ as independent variables, and $\rho V_{t+1}(X_{t+1})$ as the response variable.
The approximation function is denoted $\widehat{\Phi}_{t}$.
The method is implemented in two stages:
\begin{enumerate}
\item \textbf{Forward simulation:} For $t=0, 1, \hdots, N-1$, the random state, control, disturbance variables as well as the transitioned state are simulated as $X_{t}^{m}$, $\pi_{t}^{m}$, $Z_{t}^{m}$, and $\widetilde{X}_{t+1}^{m} = \mathcal{T}_{t}(X_{t}^{m}, \pi_{t}^{m}, Z_{t}^{m})$, $m=1,2, \hdots, M$, where $\pi_{t}$ is sampled independently from $X_{t}$.
\item \textbf{Backward recursion:} Starting from the boundary condition $V_{N}(x) = R_{N}(x)$, the optimal stochastic control problem in Equation~\eqref{eq:framework_valuefunction} is solved using the recursion in Equation~\eqref{eq:bellmanEquation}, as detailed in Algorithm~\ref{montecarlo_algorithm_realized}.
\end{enumerate}

{\color{black}\begin{remark}\label{rem:pea-connection}
The regression step to approximate \eqref{eq:conditionalexp} is closely related in spirit to regression-based expectation approximations used in computational economics, most notably the parameterized-expectations approach of \citet{den1990solving} and its neural-network implementations (e.g.\ \citet{duffy2001approximating}), as well as recent ML projection methods (e.g.\ \citet{valaitis2024machine}).
They parametrize conditional expectations that appear in \emph{equilibrium conditions} and learn them via forward in time simulation with regression.
A key difference of our approach is
that we solve a \emph{recursive stochastic control} problem backward in time by approximating the continuation value conditional expectation in the Bellman equation via regression and then recover the optimal control by a separate numerical maximization step.
This separation is convenient in stochastic DICE because controls are constrained and the state dynamics are endogenous.
\end{remark}
}

\subsection{Transformation bias and heteroskedasticity}\label{subsection-biastransform}
To mitigate challenges in approximating the value function due to the extreme curvature of utility functions, one can introduce a transformation $H(x)$ that mirrors the shape of the value function.
In our implementation, we use:
\begin{equation}\label{transform_eq}
H(x) = \frac{1}{1-\alpha} \mathrm{e}^{x(1-\alpha)}.
\end{equation}
{\color{black} For $\alpha\neq 1$, the inverse is $H^{-1}(y)=\frac{1}{1-\alpha}\log((1-\alpha)y)$, which is well-defined on the domain $\{y \in \re:(1-\alpha)y>0\}$. In particular, for $\alpha>1$ one has $H(x)<0$ for all $x$, so $H^{-1}$ is applied to negative arguments throughout.}

At each time $t<T$, the transformed value function is approximated using the least-squares regression:
\begin{equation}
H^{-1}(\rho V_{t+1}(X^{m}_{t+1})) = \bm{f}_{\theta}(X_{t}^{m}, \pi_{t}^{m}) + \epsilon_{t}^{m}, \quad m =1, 2, \hdots, M,
\end{equation}
where $\epsilon^{m}_t$, $m=1, 2, \hdots, M$ are zero mean and independent error terms, $\{ \bm{f}_{\theta} \colon \theta \in \Theta_{t} \}$ is a parametrized family of predictor functions, and $H^{-1}$ the inverse of the transformation function.
Then,
\begin{equation}\label{unbiasedestimate_eq}
\Phi_{t}(X_{t}, \pi_{t}) = \int H\big(\bm{f}_{\theta}(X_{t}, \pi_{t}) + y\big)\, \mathrm{d} F_{t}(y),
\end{equation}
where $F_{t}$ is the distribution of the error term $\epsilon_{t}$.

In the absence of a closed-form solution for the integral in Equation~\eqref{unbiasedestimate_eq}, the empirical distribution of the residuals:
\begin{equation}
\widehat{\epsilon}_{t}^{m} = H^{-1}(\rho V_{t+1}(X^{m}_{t+1})) - \bm{f}_{\theta}(X_{t}^{m}, \pi_{t}^{m})
\end{equation}
can be used to approximate this integral.
Consequently, the estimate of $\Phi_t(X_{t},\pi_{t})$ becomes:
\begin{equation}\label{eq:smearing_controlled}
\hat\Phi_{t}(X_{t}, \pi_{t}) = \frac{1}{M} \sum_{m=1}^{M} H\big(\bm{f}_{\theta}(X_{t}, \pi_{t}) + \widehat{\epsilon}_{t}^{m}\big).
\end{equation}
For the chosen transformation $H(x)$ in \eqref{transform_eq}, Equation~\eqref{eq:smearing_controlled} simplifies to:
\begin{equation}\label{eq:smearing_controlled1}
\hat\Phi_{t}(X_{t}, \pi_{t}) = H\big(\bm{f}_{\theta}(X_{t}, \pi_{t})\big)  \frac{1}{M} \sum_{m=1}^{M} \mathrm{e}^{\widehat{\epsilon}_{t}^{m} (1-\alpha)}.
\end{equation}
{\color{black}Under the homoskedasticity assumption that the distribution of $\epsilon_t$ does not depend on $(X_t,\pi_t)$, the mean of the transformed residuals in \eqref{eq:smearing_controlled1} is constant in
$(X_t,\pi_t)$ and can be precomputed and reused.}

If heteroskedasticity is present in the regression with respect to the state and control variables, a method that accounts for heteroskedasticity is required.
In this case, the conditional variance can be modeled as a function of covariates:
\begin{equation}\label{heterosced_eq}
\mathrm{var}\big(\epsilon_{t}\, |\, X_{t}; \pi_{t}\big) = \bm{g}_{\theta}(X_{t},\pi_{t}),
\end{equation}
where $\{ \bm{g}_{\theta} \colon \theta \in \widehat{\Theta}_{t} \}$ is another parametrized family of predictor functions.
There are various standard methods to estimate $\bm{g}_{\theta}$ and the \emph{smearing estimate with controlled heteroskedasticity} can then be used as discussed in \cite{andreasson2022}.

\begin{remark}\strut
\begin{itemize}
\item The method presented in Algorithm~\ref{montecarlo_algorithm_realized} is called the \emph{regression surface approach}.
A common alternative is the \emph{realized value approach}, where the value function $V_{t}(x)$ in Equation~\eqref{eq:bellmanEquation} is not computed by using the approximation of the conditional expectation (which was needed to find the optimal policy according to Equation~\eqref{optimal-policy}), but rather by computing the discounted sum of rewards along one trajectory starting from the state $x$ at time $t$.
While promising greater numerical stability than the regression surface approach, the realized value approach requires calculating optimal decisions along the individual trajectories, which comes at a significant computational cost.
For details on this approach, we refer to~\citet{andreasson2022} and references therein.
Originally, we also implemented the realized value approach, however, we found that the regression surface approach provided a sufficiently accurate solution for the number of sample points chosen in our numerical study in Section~\ref{section-numerics}.
\item Another approach worth mentioning is the \emph{regress later} LSMC method.
Here, the value function is approximated directly rather than the conditional expectation: $V_{t+1}(x) \approx \bm{f}_{\theta}(x)$.
Finding the optimal policy in \eqref{optimal-policy} then  requires the explicit calculation of the conditional expectation:
\begin{equation*}
\mathbb{E}\big[\bm{f}_{\theta}(X_{t+1})\, \big|\, X_{t} = x; \pi_{t} \big]
\end{equation*}
either analytically or numerically with quadrature methods.
However, as mentioned earlier, this approach becomes infeasible in the case of many simultaneous shocks due to the high dimensionality of the required integration.
\end{itemize}
\end{remark}

\subsection{Neural networks}\label{subsection-neuralnetworks}
In our paper, we choose for the parametrized family of functions $\{\bm{f}_{\theta} \colon \theta \in \Theta \}$ the class of deep neural networks.
This algorithmically generated class of functions has found tremendous success in all fields of science.
Over the years, it has been shown that neural networks can act as surrogate functions in many models, due to their far-reaching approximation capabilities.

{\color{black}
In classical LSMC implementations for Bermudan/American option pricing, the regression model for the continuation value is often taken to be a low-degree polynomial in the state variables \citep[e.g.][]{longstaff2001valuing,tsitsiklis2001regression,belomestny2010regression}. This can be effective when the continuation value is sufficiently smooth and the effective dimension is small. We also note that in Bermudan option pricing, the state dynamics is not affected by control and thus continuation value does not depend on control helping to get good results with a simple polynomial approximation.

In the stochastic DICE setting, the continuation value displays strong nonlinearities (utility curvature, nonlinear damages, multiplicative growth/climate components) and the mitigation control is constrained to $[0,1]$, which can produce kinked policy rules and extended boundary regions. Polynomial bases may then require higher degrees and many interaction terms to achieve a robust fit; the number of monomials of total degree at most $p$ in $d$ variables equals $\binom{d+p}{p}$ and grows rapidly with $d$ and $p$, which can lead to ill-conditioned least-squares problems and sensitivity to basis selection. Deep neural networks provide a flexible nonparametric alternative that avoids committing to an ex ante functional form and have been used as function approximators in regression-based dynamic programming methods \citep{tsitsiklis2001regression,valaitis2024machine}. Another reason for simple polynomial approximation may not work well for estimation of mitigation control is because optimal control is found in the maximisation step after conditional expectation is approximated. If the functional form of polynomial approximation wrt control is not good, then this may lead to a very bad results for optimal control. `Non-parametric' nature of neural network regression adapts to the correct shape of the function to approximate without guessing the functional form and thus should lead to better results.}

Theorems that establish approximations are referred to as \textit{universal approximation theorems} (UAT); notable contributions include \citet{cybenko89} and \citet{hornik91}.
These theorems establish the topological density of sets of neural networks in various topological spaces.
One speaks of the universal approximation property \citep{kratsios21} of a class of neural networks.
Unfortunately, these theorems are usually non-constructive.
To numerically find optimal neural networks, one typically combines backpropagation (see, for example, \citet{rumelhart86}) with ideas from stochastic approximation \citep{robbins51,kiefer52,dvoretzky56}.

Assuming sufficient integrability, the conditional expectation in Equation~\eqref{eq:conditionalexp} is the orthogonal projection of $\rho V_{t+1}(X_{t+1})$ onto the subspace spanned by $(X_{t}, \pi_{t})$ in the space of square-integrable random variables.
The universal approximation property of neural networks in this space (see, for instance, \citet[Theorem~1]{hornik91}) then justifies the approximation of $\Phi_t(X_{t},\pi_t) $ by $\bm{f}_{\theta}(X_{t}, \pi_{t})$ for a suitably chosen neural network $\bm{f}_{\theta}$.

\begin{figure}[htb!]
\centering
\begin{minipage}{.90\textwidth}
\begin{algorithm}[H]
\caption{LSMC (regression surface)}\label{montecarlo_algorithm_realized}
\begin{algorithmic}[1]
\Statex {[}Forward simulation{]}
\For{${t} = 0 \,\, \mathbf{to} \,\, N-1$}
\For{$m = 1 \,\, \mathbf{to} \,\, M$}
\State sample $X_{t}^{m}$ in the domain of its possible values \Comment{State}
\State sample $\pi_{t}^m$ in the domain of its possible values $\mathcal{A}_{t}$\Comment{Control}
\State sample $Z_{t}^m$ from the distribution specified by the model\Comment{Disturbance}
\State Compute $\widetilde{X}_{t+1}^{m} \coloneqq \mathcal{T}_{t}(X_{t}^{m}, \pi_{t}^{m}, Z_{t}^{m})$\Comment{Evolution of state}
\EndFor
\EndFor
\end{algorithmic}
\vspace{\parskip}
\begin{algorithmic}[1]
\Statex {[}Backward recursion{]}
\For{${t} = N \,\, \mathbf{to} \,\, 0$}
\If{$t = N$}
\State $\widehat{V}_{t}(\widetilde{X}_{t}) \coloneqq R_{t}(\widetilde{X}_{t})$ \label{line:terminalvalue}
\Else
\Statex $\quad \quad \quad${[}Regression of transformed value function{]}
\State $\widehat{\theta}_{t} \coloneqq \argmin_{\theta \in \Theta_{t}} \sum^M_{m=1} \left[\bm{f}_{\theta}(X_{t}^{m}, \pi_{t}^{m}) - H^{-1}(\rho \widehat{V}_{t+1}(\widetilde{X}^{m}_{t+1})) \right]^2$
\Statex $\quad \quad \quad$Approximate conditional expectation $\widehat{\Phi}_{t}(X_{t}, \pi_{t})$ using Equation~\eqref{eq:smearing_controlled}
\For{$m = 1 \,\, \mathbf{to} \,\, M$}
\Statex $\quad \quad \quad \quad$ {[}Find optimal control{]}
\State $\pi_{t}^{\ast}(\widetilde{X}_{t}^{m}) \coloneqq \argsup_{\pi_{t} \in \mathcal{A}_t} \big\{ R_{t}(\widetilde{X}_{t}^{m}, \pi_{t}) + \widehat{\Phi}_{t}(\widetilde{X}_{t}^{m}, \pi_{t}) \big\}$ \label{line:optimalcontrol}
\Statex $\quad \quad \quad \quad$ {[}Update value function{]}
\State $\widehat{V}_{t}(\widetilde{X}_{t}^{m}) \coloneqq R_{t}(\widetilde{X}_{t}^{m}, \pi_{t}^{\ast}(\widetilde{X}_{t}^{m})) + \widehat{\Phi}_{t}(\widetilde{X}_{t}^{m}, \pi_{t}^{\ast}(\widetilde{X}_{t}^{m}))$
\EndFor
\EndIf
\EndFor
\end{algorithmic}
\end{algorithm}
\end{minipage}
\end{figure}

\subsection{Uncertainty quantification}\label{uncertainty-quantification}
Uncertainty quantification (UQ) is a research field focused on understanding how uncertainties in model inputs, parameters, and other factors propagate through models to affect their outputs.
This understanding is crucial for making informed decisions based on model predictions, particularly in complex systems where such decisions can have significant consequences.
A key tool in UQ are Sobol' indices \citep{sobol01}, which are quantitative measures used in sensitivity analysis to apportion the variance of a model output to different input variables or combinations of input variables.
By identifying the most important input variables and their interactions, Sobol' indices guide efforts to sort out the main factors which should be studied with care in complex models.

{\color{black} Variance-based global sensitivity analysis based on Sobol' indices has been advocated and used in economic applications; see \citet{harenberg2019uqgsa} for a general treatment and \citet{miftakhova2021gsa} for an application to DICE. Related computational-economics work also emphasizes that surrogate/ML representations of value and policy functions can facilitate UQ in high-dimensional dynamic settings; see \citet{scheidegger2019mlhd}.}

Sobol' indices provide a comprehensive view of how input variables and their interactions influence model outputs.
They can be applied to any type of model, regardless of its complexity or the nature of its inputs and outputs.
They are particularly valuable because they capture the effects of nonlinear interactions among input variables, which is critical for understanding complex systems.
However, calculating Sobol' indices requires a large number of model evaluations, which can be computationally expensive for complex models.
The accurate estimation of Sobol' indices also depends on efficient and adequate sampling of the input space.

Denote our stochastic DICE model by $G$, which maps model inputs $X$ (such as the temperature-sensitivity coefficient) to model outputs $Y = G(X)$ (such as the projection of the global mean surface temperature in the year 2100).
There are two main types of Sobol' indices.

Denote by $U=(U_1,\dots,U_d)$ the vector of uncertain model inputs (parameters and shock drivers), assumed mutually independent, and let $Y = G(U)$ be a scalar model output of interest (e.g., $T_{2100}$ or $\mathrm{SCC}_0$). In our implementation, each $U_i$ may represent a \emph{block} of inputs (e.g., the full sequence of productivity shocks over the horizon), in which case the variance decompositions below apply verbatim with $U_i$ interpreted as that block.

\textbf{First-order Sobol' indices $S_{i}$:} These indices represent the contribution of a single input (or input block) $U_{i}$ to the output variance $\mathbb{V}(Y)$, ignoring interaction effects with other inputs:
\begin{equation*}
S_{i} = \frac{\mathbb{V}_{U_{i}}\!\big(\mathbb{E}_{U_{\sim i}}[Y\, |\, U_{i}]\big)}{\mathbb{V}(Y)},
\end{equation*}
where $\mathbb{E}_{U_{\sim i}}[Y\, |\, U_{i}]$ denotes the conditional expectation of $Y$ given $U_{i}$ with respect to all inputs $U$ except for $U_{i}$, and $\mathbb{V}_{U_{i}}(\cdot)$ denotes the variance with respect to $U_{i}$.

\textbf{Total-order Sobol' indices $S_{T_i}$:} These indices represent the contribution of an input variable to the output variance, including all interactions with other variables.
They are defined as:
\begin{equation*}
S_{T_{i}} = 1 - \frac{\mathbb{V}_{U_{\sim i}}\!\big(\mathbb{E}_{U_{i}}[Y\, |\, U_{\sim i}]\big)}{\mathbb{V}(Y)} .
\end{equation*}
where $\mathbb{E}_{U_{i}}[Y\, |\, U_{\sim i}]$ denotes the conditional expectation of $Y$ with respect to $U_{i}$ given all inputs $U$ except for $U_{i}$, and $\mathbb{V}_{U_{\sim i}}(\cdot)$ denotes the variance with respect to all inputs $U$ except for $U_{i}$.

First- and total-order Sobol' indices help determine which input variables are the most influential.
Variables with high first-order indices have a strong direct effect, while those with high total-order indices are significant due to their interactions with other variables.
In Section \ref{section-numerics}, we will compute Sobol' indices for our five identified uncertainties and examine their effect on the most important model parameters.
It is important to note that computing Sobol' indices in conjunction with the LSMC method involves solving the model with the backwards recursion~\eqref{eq:bellmanEquation} only once, and then generating a sufficiently large amount of forward trajectories to estimate the indices $S_{i}$ and $S_{T_i}$.

\subsection{Comparison with other methods}\label{subsection-comparison}
\citet{jensen2014} analyze long-term economic growth uncertainty in a DICE based assessment model with an infinite-horizon.
They express uncertainty in terms of stochastic shocks to the growth rate of total factor productivity.
The value function is approximated by Chebyshev polynomials, and the system is solved by value function iteration.
The base model has only 3 physical state variables: capital $K_{t}$, atmospheric carbon $M_{t}$, and technology level $A_{t}$.

\citet{nordhaus2018} considers the same DICE model version as the one used in this paper.
Five uncertainties are identified, the same as those explained in Subsection~\ref{subsection-uncertainty}.
These uncertainties are treated as initial parameter uncertainties.
The distributions are discretized to reduce the computational burden, thereby reducing the number of possible scenarios from an uncountably infinite amount to just a few thousands.
A Monte-Carlo-based parameter perturbation analysis is performed, where parameters are sampled, and then the corresponding deterministic version of the DICE model is solved.
In contrast to \citet{nordhaus2018}, we don't need to discretize the distributions, and we need to solve the model only once.

\citet{cai2019} also study a stochastic version of the DICE model, extending the deterministic 6-dimensional model to a stochastic 9-dimensional model.
Two additional model dimensions are due to uncertainty in the evolution of total factor productivity, and one additional dimension is due to a stochastic tipping point process.
The stochastic processes are discretized, and the resulting model is solved by value function iteration, where the value function is approximated by Chebyshev polynomials.
The model is solved with the Blue Waters supercomputer, using 110,688 cores in parallel, with computation times of up to 8 hours.
While we do not include a tipping point process in this paper, our simulation based method drastically reduces the computational burden by solving our 11-dimensional (in contrast to the 9-dimensional version of \citet{cai2019}) model formulation on a 64 core machine within around 18 hours of computation time, depending on the amount of numerical precision that is required for the solutions.
{\color{black} Expressed in terms of pure core-hours (cores $\times$ wall-clock hours) - a coarse normalization and not an apples-to-apples performance metric across different model variants - this corresponds to a reduction of more than $99\%$ relative to the implementation reported in \citet{cai2019}.}

\citet{ikefuji2020expected} formulate a stochastic version of the DICE model considering one uncertainty at a time: a) uncertainty in the damage-abatement fraction, b) uncertainty in the damage parameter, c) uncertainty in the emissions-to-output ratio, and d) uncertainty in total factor productivity $A_{t}$.
These uncertainties are introduced by multiplying the corresponding deterministic DICE variables by stochastic disturbances.
Thus, the number of state variables is the same as in the deterministic DICE (6).
To the best of our knowledge, this is the only attempt to solve a stochastic version of the DICE model by using an LSMC type approach.
They use least-squares regression with polynomial basis functions to approximate the value function, i.e.\ in the spirit of regress later LSMC.
Here, we note that their regression type Monte Carlo algorithm setup omits the integration for the conditional expectation in the Bellman equation, assuming the random disturbance is known in the transition of state variables.
In principle, the standard regress later LSMC can be implemented here to handle this type of uncertainty but it will be a subject of the curse of dimensionality in the case of more than one shock.

\citet{friedl2023deep} present a method for solving integrated assessment models and performing uncertainty quantification.
They exemplify their approach on a version of the DICE model with uncertainties in equilibrium climate sensitivity (that contains a Bayesian learning component), and the damage function (represented by a stochastic tipping process).
First, a deep neural network is trained to output, in particular, the optimal policies and value function at a given point in time, and then a Gaussian process-based model is trained to approximate quantities of interest such as the social cost of carbon in order to speed up the evaluation when calculating UQ metrics.
In contrast to \citet{friedl2023deep}, our method approximates the conditional expectation rather than the policy functions, and then finds those by running an optimizer to solve Equation~\eqref{optimal-policy}.
Approximating $\mu_{t}$ by a regression scheme is a challenging task, since the presence of the bounds (i.e.\ $0 \le \mu_{t} \le 1$) require a very careful choice of an appropriate regression scheme that can effectively interpolate the optimal policy, especially in the presence of extended periods when the policy is on the boundary.
{\color{black}Our approach avoids this issue by computing the optimal policy via constrained numerical optimization. Once the continuation-value conditional expectation has been approximated, the per-state optimization can be solved to high numerical precision; overall wall-clock time, however, is dominated by the regression step (see Section~\ref{section-numerics} and Table~\ref{tab:benchmarking_methods}).
}
Moreover, the deep LSMC method requires performing a least-squares regression, where the loss function is the squared distance between the object of interest and the neural network prediction.
This choice of loss function is significantly simpler, as it avoids the eleven individual components that enter the loss function based on an elaborate set of first-order conditions that are needed in the solution of \citet{friedl2023deep}.
Finally, in contrast to \citet{friedl2023deep}, we find that there is no need to train an additional Gaussian process-based surrogate model to perform UQ for the quantities of interest (such as the social cost of carbon).
Once the backward recursion (Equation~\eqref{eq:bellmanEquation}) has been performed, a large amount of optimal trajectories for different realizations of uncertainties can be computed easily in order to perform UQ for the quantities of interest.

{\color{black} %
Cross-paper performance comparisons are inevitably imperfect because model structures, uncertainty blocks, and numerical targets differ across contributions. To make the trade-offs assessable on a transparent yardstick, we therefore report core-hours (cores $\times$ hours) as a simple hardware-normalized summary. Table~\ref{tab:benchmarking_methods} summarizes the main computational features of the representative approaches discussed in this subsection.

\begin{table}[htbp]
\small
\centering
\begin{threeparttable}
\caption{Indicative benchmarking across representative stochastic DICE solution approaches}\label{tab:benchmarking_methods}
\begin{tabular} {@{}p{2.7cm}p{3.2cm}cp{4.2cm}p{2.2cm}@{}}
\toprule
\textbf{Reference} & \textbf{Numerical algorithm} & \textbf{\#states} & \textbf{Uncertainty structure} & \textbf{Core-hours} \\
\midrule
\citet{cai2019} & polynomial VFI, discretized shocks & 9 & productivity shock, tipping & $\approx 8.9 \cdot 10^{5}$ \\
This paper & deep LSMC, no discretization & 11 & 3 parameter shocks, \hspace{1.cm} 2 process shocks & $\approx 1.2 \cdot 10^{3}$ \\
\citet{friedl2023deep} & DEQN, first-order equilibrium & 16 & 6 parameter shocks, tipping, Bayesian learning & $\approx 3.2 \cdot 10^{1}$ \\
\bottomrule
\end{tabular}
\begin{tablenotes}[para]\footnotesize
Reported runtimes are taken from the cited papers and from our implementation details in Section~\ref{section-numerics}. Core-hours are defined as (number of CPU cores)$\times$(wall-clock hours) and are intended as a coarse normalization only.
\end{tablenotes}
\end{threeparttable}
\end{table}

In our baseline specification, the full backward recursion (including training the deep neural network regressors at each time step) takes between 9 and 18 hours on a 64-core CPU machine, corresponding to approximately $6\cdot 10^2$--$1.2\cdot 10^3$ core-hours (see Section~\ref{section-numerics} for details); subsequent uncertainty-quantification outputs are then generated via forward simulation.

The practical choice between deterministic grid/VFI approaches and simulation-based LSMC depends primarily on state dimension, the nature of uncertainty, and the intended UQ outputs. Deterministic grid methods (e.g., Chebyshev-based VFI with discretized shocks) can be attractive and highly accurate in low-dimensional settings with few shocks; however, they typically require explicit shock discretization and their computational cost grows rapidly with dimension. Simulation-based LSMC becomes preferable when (i) the state is moderately to highly multi-dimensional, (ii) uncertainty enters through multiple continuous shocks and/or parameter uncertainty without natural discretization, and (iii) one aims to produce UQ outputs (e.g., global sensitivity analysis) that require many forward evaluations. In such settings, deep neural-network regressors offer a numerically stable alternative to polynomial bases whose complexity grows quickly with dimension and interaction order (see also Section~\ref{subsection-neuralnetworks}). Conversely, in small-state benchmark problems, classical VFI/projection methods or low-order polynomial LSMC may be simpler and computationally preferable.
}

\section{Numerical study}\label{section-numerics}
In this section, we present the numerical results from applying the least-squares Monte Carlo method with transformation bias adjustment and neural network approximation of conditional expectations.
For clarity, we emphasize that our state vector $X_{t}$ consists of 11 variables: the six variables from the deterministic formulation of the DICE model ($K_{t}$, $M_{t}$, $T_{t}$), the two stochastic processes $A_{t}$ and $\sigma_{t}$, as well as the three parameters discussed in Subsection~\ref{subsection-uncertainty} (temperature-sensitivity coefficient, damage coefficient and carbon cycle coefficient).

\begin{figure}[htb!]
\centering
\captionsetup{width=0.75\textwidth,format=plain}
\includegraphics[width=0.75\linewidth]{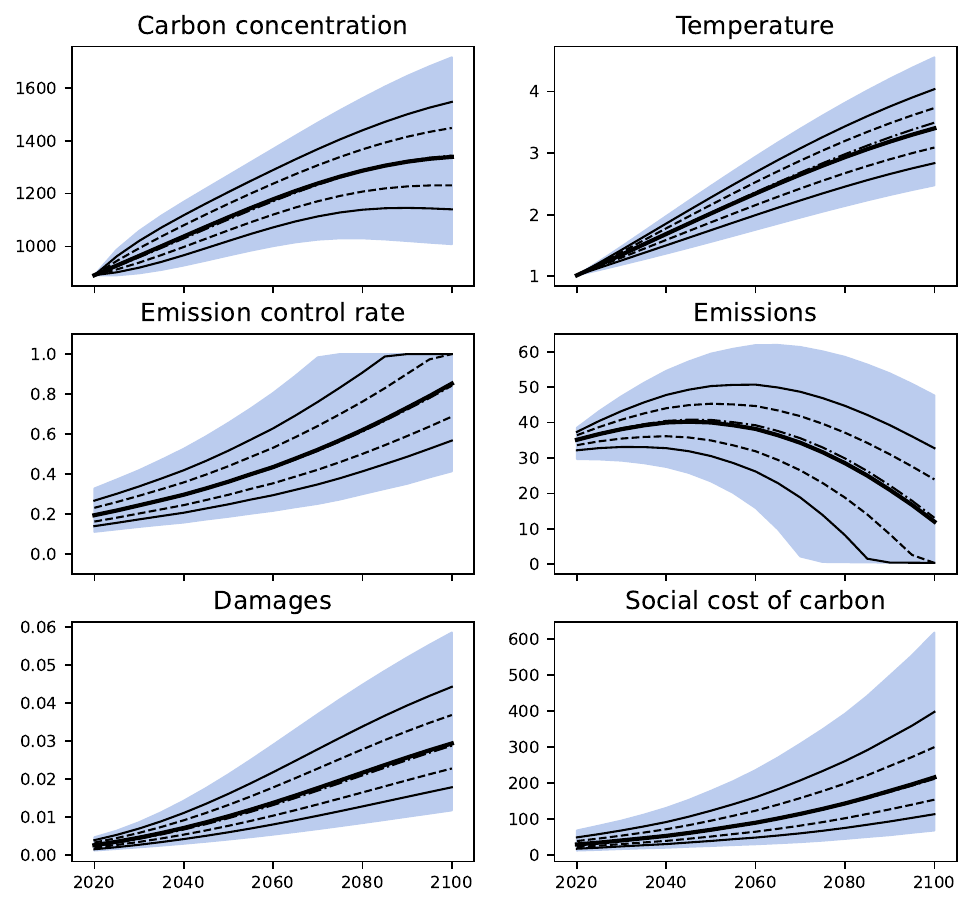}
\vspace*{-0.40cm}
\caption{Evolution of the six most important variables over time. {\color{black} The deterministic DICE-2016R2 trajectories are omitted because they are nearly indistinguishable from the plotted best-guess paths and would overlap visually.}
}
\label{fig:panel_A}
\end{figure}

{\color{black}For the backward recursion and least-squares approximation of the value function, we use approximately $8.4$ million ($2^{23}$) sample points in the 11-dimensional state space. Figure~\ref{fig:panel_A} is based on $5 \times 10^{5}$ forward trajectories, while the statistics reported in Table~\ref{tab:statistics} are based on a sample of size $10^{6}$.
To find the optimal policies in \eqref{optimal-policy}, we use the limited-memory Broyden--Fletcher--Goldfarb--Shanno algorithm with box constraints (L-BFGS-B).
On a 64 core machine, it took between 9 hours (for $\approx 4.2$ million, i.e.\ $2^{22}$, samples) and 18 hours (for $\approx 8.4$ million, i.e.\ $2^{23}$, samples) to perform the backward recursion.}
Computing optimal forward trajectories then typically took around 15 minutes for $10^5$ trajectories, and 1 hour for $5 \times 10^{5}$ trajectories.

\begin{figure}[htb!]
\centering
\captionsetup{width=0.75\textwidth,format=plain}
\includegraphics[width=0.75\linewidth]{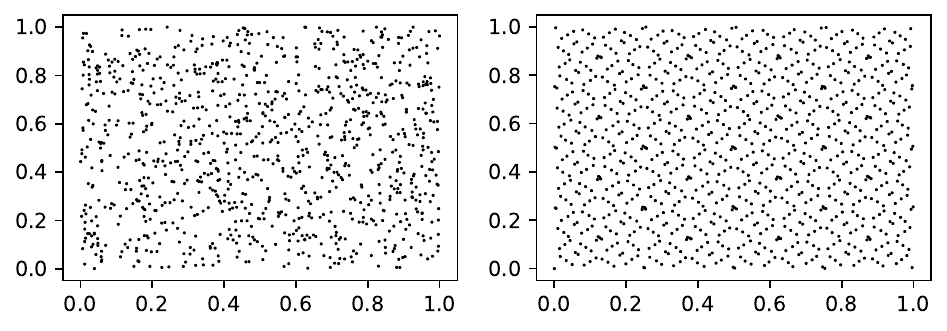}
\vspace*{-0.40cm}
\caption{Comparison of uniform grid (left panel) and low-discrepancy Sobol grid (right panel). In both cases, 1024 points were drawn in 11 dimensions. The plots depict the point distributions from the 11-dimensional grid projected on the first two components.}
\label{ld-grid}
\end{figure}

{\color{black}The numerical stability and precision of the deep LSMC recursion in this application relied primarily on three ingredients: (i) a post-decision state formulation (Eq.~\eqref{equation-post-decision}), which reduces the effective dimension of the regression for conditional expectations; (ii) normalization of rapidly growing economic variables (capital and productivity) in effective-labor units; and (iii) low-discrepancy Sobol' sequences for sampling the high-dimensional state space (Fig.~\ref{ld-grid}). We briefly emphasize these choices here to facilitate reproducibility and to clarify which components helped reduce the computational burden in our setting.}

{\color{black}The initial year for the version of the DICE model used in \citet{nordhaus2018} is 2015 (i.e., $t=0$ under $\Delta=5$-year steps). Since our numerical discussion and UQ outputs focus on the twenty-first century, we report forward trajectories from 2020 onward (i.e., $t\ge 1$). For this purpose, we anchor the initial 2015--2020 step to the deterministic DICE benchmark and then apply the stochastic optimal policy from $t\ge 1$ under the full uncertainty specification. This convention affects only the initial step and is used solely as an initialization choice; it does not alter the deep LSMC backward recursion.}
Moreover, the original DICE model is formulated as an infinite-horizon control problem, see Equation~\eqref{eq:baseDICE}.
However, our formulation of the LSMC method as discussed in Section~\ref{section-LSMC} assumes a finite time horizon with $N$ time steps ($N=97$ in our case corresponding to $t=0$ being the year 2015, and $t=N$ being the year 2500).
Imposing a finite time horizon corresponds to a truncation of the problem, and one needs to choose an appropriate boundary reward function $R_{N}(x)$.
{\color{black}This long-horizon finite truncation is standard in numerical implementations of DICE-style models: in practice (including Nordhaus-style implementations), the infinite-horizon objective is evaluated on a sufficiently long but finite horizon because discounting and convergence toward a long-run regime render further horizon extension immaterial for the welfare-relevant quantities in the twenty-first century.}
Similarly, as in \citet{cai2019}, our terminal reward function is computed by assuming that after 2500 the policies are fixed to $\mu_{t}=1$, $c_{t} = 0.78$, and that the system evolves deterministically.
The reward is then equal to the discounted sum of population-weighted utility of per-capita consumption from following the fixed policies for another 100 time steps.
Due to discounting and the large amount of time steps, it is assumed that a different choice of boundary reward that far ahead in the future should have a negligible impact on results for the twenty-first century.

For approximating conditional expectations, we use deep feedforward neural networks with two hidden layers, each containing 32 hidden nodes with hyperbolic tangent (tanh) as activation function, and a linear readout in the output layer.
Neural network training is performed using minibatch stochastic gradient descent with the Adam optimizer.
The initial learning rate is set to $10^{-3}$ and reduced to a minimum of $10^{-5}$ during training.
Early stopping is implemented to avoid overfitting.
During the backward recursion, the trained neural network from one step (e.g. step $t+1$) is used as the initial neural network for the next step's training (step $t$), which reduces computation time.

For this version of the stochastic DICE model, the transition equation~\eqref{eq:transitionfunction} can be separated into two transitions:
\begin{equation}\label{equation-post-decision}
X_{t+1} = \widetilde{\mathcal{T}}_t(F(X_t, \pi_t), Z_{t}),
\end{equation}
where the deterministic transition to the \emph{post-decision} variable $\hat{X}_{t} = F(X_{t}, \pi_{t})$ precedes the transition $X_{t+1} = \widetilde{\mathcal{T}}_t(\hat{X}_{t}, Z_{t})$.
This allows the conditional expectation in \eqref{eq:bellmanEquation} to be simplified to:
\begin{equation*}
\mathbb{E}\big[\rho V_{t+1}(X_{t+1})\, \big|\, X_{t}; \pi_{t} \big] = \mathbb{E}\big[\rho V_{t+1}(X_{t+1})\, \big|\, \hat{X}_{t} \big] .
\end{equation*}
This method offers two main advantages: (1) dimension reduction in the covariates needed for the least-squares approximation of the conditional expectation, and (2) an increase in sampling efficiency by sampling only the post-decision states $\hat{X}_{t}$ rather than both $X_{t}$ and $\pi_{t}$.
Our method benefits significantly from using post-decision variables, and we found a notable improvement in numerical precision.

Economic capital $K_{t}$ and total factor productivity $A_{t}$ can grow quite rapidly over time, especially in scenarios where large growth in $A_{t}$ meets a low consumption rate $c_{t}$.
This poses an important numerical challenge, since an appropriate domain for sampling the state variables needs to be chosen with care.
A popular solution to this issue, having been applied successfully in \citet{jensen2014}, is to normalize economic capital as follows.
First, we re-write output to express it in terms of labor-augmenting technology: $Y_{t} = A_{t} K_{t}^{\gamma} L_{t}^{1-\gamma} = K_{t}^{\gamma} (\widetilde{A}_{t}L_{t})^{1-\gamma}$, where $\widetilde{A}_{t} = A_{t}^{1/(1-\gamma)}$.
Let $\widetilde{A}_{t}^{\mathrm{det}}$ denote the deterministic trajectory of $\widetilde{A}_{t}$, where $g_{A}(t)$ is fixed to be equal to the expected value.
Economic capital and output are then expressed in terms of units of effective labor: $k_{t} = K_{t} / (\widetilde{A}_{t}^{\mathrm{det}}L_{t})$, and $y_{t} = Y_{t} / (\widetilde{A}_{t}^{\mathrm{det}}L_{t})$.
The state variable $A_{t}$ can also be substituted by $\widetilde{A}_{t}$ and further normalized to $a_{t} = \widetilde{A}_{t} / \widetilde{A}_{t}^{\mathrm{det}}$.
In our simulations, we found that these normalization steps had a favorable impact on the precision of the numerical results.

Calculating the social cost of carbon \eqref{equation-SCC} requires knowledge of partial derivatives of the value function with respect to atmospheric carbon concentration and economic capital.
Since we do not have an analytic representation of the value function, we follow an approximation approach that was discussed in \citet{traeger2014}, where Chebyshev polynomials were used to approximate the value function.
At each time $t$, we approximate the value function $V_{t}$ by a neural network:
\begin{equation}\label{equation-VFapprox}
V_{t}(x) \approx \bm{g}_{\theta}(x),
\end{equation}
for a suitable parameter vector $\theta$.
This approach strikes a balance between numerical precision and analytical tractability, applicable even in the presence of stochasticity.
Note that the idea of approximating the value function by a neural network has already been carried out in \cite{kelly2001solving} where, however, the neural network approximation was not used for computing the social cost of carbon.

The post-decision variables $\hat{X}_{t}$, representing the states $X_{t}$ after decision $\pi_{t}$, have the same dimension as $X_{t}$.
The sampling step in Algorithm~\ref{montecarlo_algorithm_realized} requires choosing an effective sampling distribution.
One standard approach would be to put a high-dimensional grid of uniformly drawn points around the deterministic DICE solution.
However, in order to improve numerical precision, low-discrepancy grids are favorable in order to keep the number of sample points needed to a reasonable amount.
Latin hypercube sampling offers a more favorable distribution of grid points compared to uniform sampling.
We chose to use Sobol' grid points \citep{sobol67}, which offer even higher numerical precision compared to Latin hypercube samples.
Figure \ref{ld-grid} shows the point distribution of a uniform and of a Sobol' grid for comparison.
We found that using a low-discrepancy grid improved the numerical precision of the results.

\begin{figure}[htb!]
\centering
\captionsetup{width=0.75\textwidth,format=plain}
\includegraphics[width=0.6\linewidth]{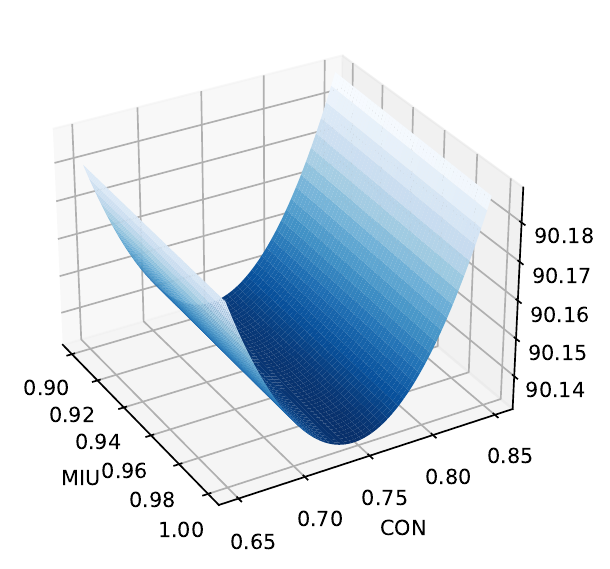}
\vspace*{-0.40cm}
\caption{Typical optimization surface over $(c_{t}, \mu_{t})$ encountered during backward recursion.}
\label{optim-surface}
\end{figure}

A major challenge in solving the model was to obtain stable estimates of the optimal emission mitigation rate $\mu_{t}$.
Estimating the optimal consumption rate $c_{t}$ was straightforward, but estimating $\mu_{t}$ required very precise estimates in the least-squares approximation of the conditional expectation.
Figure~\ref{optim-surface} offers a partial explanation.
It illustrates a typical optimization surface when trying to find the optimal policies $(c_{t}, \mu_{t})$ in Equation~\eqref{optimal-policy}, showing a steep curvature for $c_{t}$ and a much flatter surface for $\mu_{t}$, indicating the need for precise numerical approximations and small tolerance values in the optimizer.
We see this issue as a consequence of the model setup.
For example, a low carbon intensity $\sigma_{t}$ for times after $2100$ leads to low emissions and mitigation costs, resulting in an almost negligible effect of the mitigation rate on the value $V_{t}$.
In order to resolve this issue, very precise numerical approximations of conditional expectations based on a large number of well-spaced sample points as well as small tolerance values in the optimizer for $(c_{t}, \mu_{t})$ were required.

Each point $x$ in the state space can be optimized independently in Equation~\eqref{optimal-policy}.
In other words, when solving \eqref{optimal-policy} over a high-dimensional grid in state space, the individual optimization steps for each grid point can be executed in parallel.
This parallel optimization is implemented using Python's \texttt{multiprocessing} package over 64 cores, significantly reducing computation time and allowing for the usage of a reasonably large sample size without excessive computational costs.

Figure~\ref{fig:panel_A} presents the evolution of the six most important variables over time if the optimal strategy is used, based on 500,000 independently simulated trajectories.
These six variables are the social cost of carbon $SCC_t$, the global mean surface temperature $T_t^{\mathrm{AT}}$, the carbon concentration in the atmosphere $M_t^{\mathrm{AT}}$, the emission mitigation rate $\mu_{t}$, total $\mathrm{CO}_{2}$ emissions $E_{t}$, and damages $\pi_{2} \times (T_{t}^{\mathrm{AT}})^{2}$.
The panels include the median trajectory (bold solid line), expected trajectory (dash-dotted line), the 25$\%$ and 75$\%$ quantiles (dashed lines), the 10$\%$ and 90$\%$ quantiles (solid lines) as well as the range of sampling paths between the 1$\%$ and 99$\%$ quantiles (shaded area).
We can observe a significant amount of uncertainty in all variables.
Most notably, a significant fraction of scenarios sees full mitigation (i.e.\ $\mu_{t} = 1$) well before the year 2100 in the optimal case, though the median trajectory is a bit below the full mitigation in 2100.
We also observe that for temperature, the 1$\%$ quantile is approximately at 2.5$^\circ$C, while the 99$\%$ quantile is approximately at 4.5$^\circ$C.
The SCC is about US\$200 in 2100 under the median trajectory, and between \$150 and \$300 for the 10\% and 90\% quantiles.
{\color{black} For completeness, we compared the evolution of all controls, state variables and derived quantities in the best-guess (mean-parameter) case to the deterministic DICE-2016R2 solution. The resulting trajectories are nearly indistinguishable. We therefore do not print the deterministic DICE-2016R2 curves in Figure~\ref{fig:panel_A}, since the lines would overlap and cannot be visually separated.}

\begin{figure}[htb!]
\centering
\captionsetup{width=0.75\textwidth,format=plain}
\includegraphics[width=0.75\linewidth]{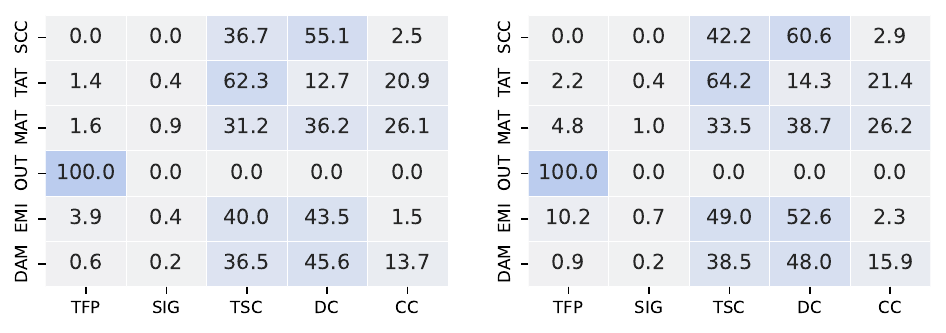}
\vspace*{-0.40cm}
\caption{First-order (left) and total-order (right) Sobol' indices for various model outputs with respect to uncertainty in total factor productivity (TFP), carbon intensity (SIG), temperature-sensitivity coefficient (TSC), damage coefficient (DC) and carbon cycle coefficient (CC).}
\label{fig:panel_B}
\end{figure}

Figure~\ref{fig:panel_B} shows the first- and total-order Sobol' indices for various model outputs in relation to the 5 sources of uncertainty which we considered in the model.
The analyzed outputs are the social cost of carbon in 2020 (SCC), the mean surface temperature in the atmosphere in 2100 (TATM), the carbon concentration in the atmosphere in 2100 (MAT), output in 2100 (OUT), emissions in 2100 (EMI) as well as damages in 2100 (DAM).
The first-order Sobol' indices (left panel) illustrate the individual contribution of each input to the variance of the outputs, while the total-order Sobol' indices (right panel) capture the overall contribution, including interactions with other inputs.
Note that first-order indices do not sum up to $100\%$, as we have not taken into account higher order indices (second order, third order etc.).

{\color{black} Sobol' indices quantify the contribution of each uncertain input to the \emph{variance} of the output and are therefore non-directional. To make the sign of the overall impact of uncertainty transparent, Table~\ref{tab:statistics} compares the best-guess value to the mean, providing a simple summary of the ``uncertainty premium'' for each reported quantity.
}

\begin{figure}[htb!]
\centering
\captionsetup{width=0.75\textwidth,format=plain}
\includegraphics[width=0.75\linewidth]{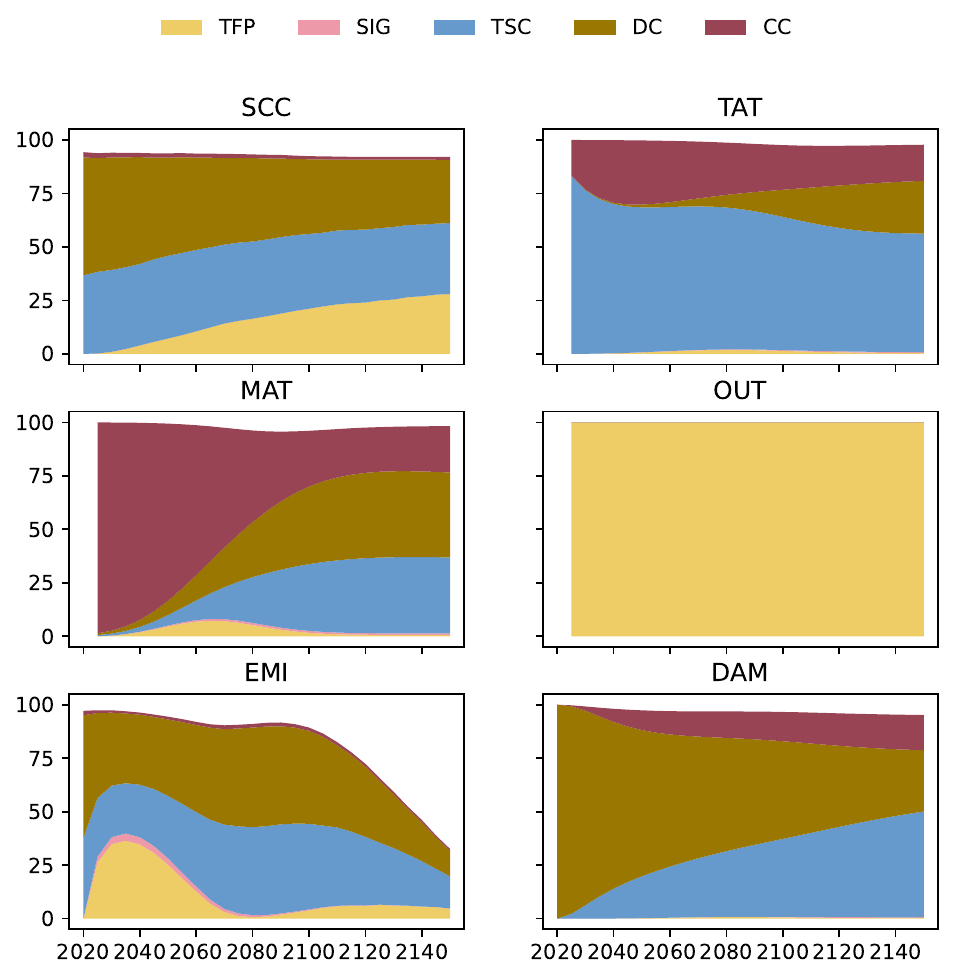}
\vspace*{-0.40cm}
\caption{First-order Sobol' indices for main variables over time.}
\label{fig:panel_C}
\end{figure}

{\color{black}
\begin{remark}
A deterministic DICE benchmark is studied by \citet{miftakhova2021gsa}, who performs a broad parametric global sensitivity analysis using a surrogate (polynomial-chaos) representation and reports Sobol' indices for quantities such as the SCC and optimal policy. While our analysis differs in that (i) we solve a recursive stochastic-control formulation with process shocks and endogenous state-dependent policy feedback and (ii) we restrict attention to five benchmark uncertainty blocks (from \citet{nordhaus2018}) rather than the full DICE parameter set, the qualitative messages for overlapping drivers are aligned: productivity uncertainty dominates output variability, whereas climate-related parameters (temperature sensitivity and damages) dominate temperature/damage variability and are important for SCC-related uncertainty. Moreover, the gap between first- and total-order indices (Figure~\ref{fig:panel_B}) and the time variation in first-order indices (Figure~\ref{fig:panel_C}) highlight interaction effects and time dependence that naturally arise under an endogenous optimal policy response in the stochastic formulation.
\end{remark}
}

{\color{black} %
From Figure~\ref{fig:panel_B}, output is predominantly impacted by total factor productivity, with both first-order and total-order indices around 100\%, reflecting the direct Cobb--Douglas link between productivity and gross output. Productivity shocks also affect emissions and thus climate outcomes mechanically through higher output and industrial emissions; however, conditional on the benchmark uncertainty ranges and the endogenous optimal policy response, their contribution to the variance of temperature- and damage-related outputs at the reported time horizons is small relative to climate-module parameters (see the Sobol' shares for $T^{AT}_{2100}$ and damages in Figure~\ref{fig:panel_B}).

In contrast, within the benchmark uncertainty specification adopted from \citet{nordhaus2018}, the contribution of carbon-intensity (decarbonization) uncertainty to the variance of the selected outputs is below 1\% over the time horizons reported in Figure~\ref{fig:panel_B}. This statement is meant in the variance-decomposition sense and conditional on the remaining uncertainties and on the endogenous optimal policy response; in particular, a substantial fraction of optimal-policy scenarios reaches (near) full mitigation by the end of the century, which mechanically attenuates the sensitivity of late-horizon industrial emissions to the emissions-intensity path. Accordingly, this finding should not be interpreted as a claim that \emph{recalibrated} emissions-intensity pathways (e.g.\ SSP/RCP-based trajectories) are quantitatively unimportant; such scenario analysis is a complementary ``between-scenario'' layer \citep{oneill2016scenariomip,riahi2017ssp,murakami2025scenario}.

Overall, Figure~\ref{fig:panel_B} highlights that:
\begin{enumerate}
\item Productivity uncertainty dominates the variance of output in our benchmark set-up.
\item Carbon-intensity (decarbonization-rate) uncertainty contributes comparatively little to the variance of the
reported quantities in this benchmark.
\item Climate-related parameters (climate sensitivity and the damage coefficient) account for most of the
uncertainty in temperature-, damage-, and SCC-related outputs, consistent with existing DICE sensitivity analyses (see Remark~4.1 and \citealp{miftakhova2021gsa}).
\end{enumerate}
}



Figure~\ref{fig:panel_C} shows the evolution of first-order Sobol' indices for our main variables over time, up to the year 2150.
It highlights the fact that the impact of the uncertain variables on the outputs changes over time.
Most notably, the changes appear not to follow a linear pattern, especially when looking at emissions.
There, the impact of total factor productivity $A_{t}$ peaks around the year 2035, but declines rapidly afterwards.
In contrast, the impact of $A_{t}$ on the social cost of carbon gradually rises from 0$\%$ in the year 2020, to around 25$\%$ in the year 2150.
This does not come as a surprise, as it highlights the effect of the large initial uncertainty about parameters such as the temperature-sensitivity and damage coefficients, which combines with a negligible initial uncertainty in total factor productivity that grows over time.
Another interesting effect that can be observed is that the total sum of all first-order indices declines for emissions from above 95$\%$ in the year 2020 to slightly below 40$\%$ in the year 2150.
This motivates the insight that the impact due to interactions between the uncertain variables grows over time.

Table~\ref{tab:statistics} shows the key statistics for the major variables.
In terms of the coefficient of variation (CV), we can observe the highest degree of uncertainty in emissions, followed by the social cost of carbon, damages, and output.
Most importantly, the interquartile range (IQR) of 0.64$\degree$C for temperature and 1.4$\%$ for damages highlights the importance of considering the notable variations in projections due to the presence of uncertainty.
Moreover, we can re-confirm the presence of noticeable differences between the mean, median and best guess values for some variables, which is in line with the observations of \citet{nordhaus2018}.
Differences between the mean and median values hint at the presence of skewness in the distribution of the variables, which can also be visually confirmed from Figure~\ref{fig:panel_A}.
Finally, differences between the best guess estimates and the mean and median values show that in some cases, the best guess provides a reasonable approximation of the complex dynamics, whereas in other cases it does not, which again highlights the importance of explicitly including stochastic dynamics into climate-economy models.

\begin{table}[htb!]
\small
\centering
\begin{threeparttable}
\caption{Statistics for major variables}\label{tab:statistics}
\begin{tabular}{@{}lccccccc@{}}
\toprule
\textbf{Variable} & \textbf{Mean} & \textbf{BG} & \textbf{Median} & \textbf{SD} & \textbf{IQR} & \textbf{CV} & \textbf{Mean-BG}\\
\midrule
Social cost of carbon, 2020 & 30.9 & 28.3 & 28.7 & 12.5 & 16.7 & 0.40 & +9.2 \\
Temperature, 2100 ($\degree$C) & 3.42 & 3.49 & 3.40 & 0.46 & 0.64 & 0.13 & $-2.0$ \\
Carbon concentration, 2100 (ppm) & 1,342 & 1,344 & 1,339 & 156 & 217 & 0.12 & $-0.1$ \\
World output, 2100 (trillions, 2015$\$$) & 833.6 & 795.9 & 811.2 & 203.6 & 271.9 & 0.24 & +4.7 \\
Emissions, 2100 & 14.0 & 13.1 & 12.0 & 13.3 & 23.6 & 0.95 & +6.9 \\
Damages, 2100 (percent output) & 3.0 & 2.9 & 2.9 & 1.0 & 1.4 & 0.34 & +3.4 \\
\bottomrule
\end{tabular}
\begin{tablenotes}[para]\footnotesize
SD, IQR and CV refer to standard deviation, interquartile range and coefficient of variation, respectively.
BG refers to best guess, which is the value calculated along the expected trajectory, assuming that uncertainties are set to their respective means.
{\color{black}Mean-BG is computed as $100\times(\text{Mean}/\text{BG}-1)$ and provides the sign of the overall impact of uncertainty on each quantity (in $\%$).}
\end{tablenotes}
\end{threeparttable}
\end{table}

{\color{black} %
These comparisons show that, in our benchmark CRRA-DICE specification, uncertainty has a modest effect on the \emph{typical} (median) optimal-policy trajectory, but it can materially affect \emph{derived} quantities such as the SCC through distributional asymmetry. In particular, the SCC is right-skewed at the initial date, so the mean SCC exceeds the median value. The sign and magnitude of this effect is model- and preference-specification dependent, which helps reconcile differing findings in the literature (e.g.\ \citealp{cai2019,benmir2020green}).
}

{\color{black} %
One reason the benchmark uncertainty effects are modest for median policy trajectories is that the standard CRRA specification links risk aversion and intertemporal substitution. More risk-sensitive recursive preferences that separate IES and risk aversion (e.g.\ Epstein--Zin; see \citealp{epstein1989substitution,weil1990nonexpected} and the IAM implementation in \citealp{cai2019}) can strengthen precautionary incentives and thereby amplify the effect of uncertainty on the SCC and mitigation. While we keep CRRA in the benchmark calibration to retain comparability with DICE-2016R2, our numerical framework accommodates these preference specifications with only a modification of the continuation-value recursion and regression target.
}

\section{Conclusions}\label{section-conclusion}
Climate-economy models are essential tools for informed decision-making, risk management, and strategic planning in the face of climate change.
These models provide a structured framework for analyzing the economic implications of climate policies and developing sustainable solutions to mitigate and adapt to climate change impacts.
Incorporating stochastic models into climate-economy analyses is crucial for capturing the full spectrum of uncertainties, improving risk assessment, designing resilient policies, and enhancing the overall robustness and reliability of the models and their predictions.
However, the complexity of capturing the intricate, multifaceted, and probabilistic nature of climate and economic systems, coupled with the computational challenges of handling large-scale, high-dimensional, and stochastic models, poses significant challenges in deriving efficient solutions in the presence of uncertainty.

This paper presents an advanced approach to modeling recursive stochastic climate-economy models using a deep least-squares Monte Carlo (LSMC) method.
The method's flexibility allows for the application to various types of uncertainties, including parametric and stochastic process uncertainties.
The integration of deep neural networks enables the handling of high-dimensional models in a tractable manner and within a reasonable computational budget, thus making stochastic climate-economy models more accessible to researchers and policymakers.
The methodology and findings presented here provide a solid foundation for future work in this vital area of research.
{\color{black}More broadly, the resulting scenario distributions can serve as inputs for downstream climate-finance and actuarial applications; see, e.g., \citet{arandjelovicShevchenko2025netzero,arandjelovicShevchenko2025reserves}.}

Future research should explore the incorporation of Bayesian learning mechanisms to update probabilities as more information becomes available over time.
Since our approach can manage high-dimensional stochastic shocks, a natural next step is to study the impact of multi-dimensional probability distributions whose marginals are correlated.
{\color{black} Another natural extension is to incorporate risk-sensitive recursive preferences that separate IES and risk aversion (e.g.\ Epstein--Zin), as in \citet{cai2019}, to quantify how preference specification alters precautionary motives and the SCC under the same uncertainty blocks.}
Additionally, we aim to apply our method to the study of climate tipping points as well as the Regional Integrated model of Climate Change and the Economy (RICE) of \citet{nordhaus1996}.
These future steps could further refine the model's predictions and enhance its policy relevance.

{\color{black}
Extending from the global DICE model to the multi-region RICE model \citep{nordhaus1996} increases the dimension of the economic state (e.g.\ regional capital stocks) and the number of controls (regional consumption and mitigation). In the deep LSMC algorithm, however, the regression step targets the continuation-value conditional expectation as a function of the post-decision state; hence the effective regression input dimension scales with the number of endogenous state variables rather than with the number of controls. The dominant costs are (i) simulating training samples and (ii) fitting the regression model at each time step, both of which are parallel and can be distributed across CPU/GPU resources. While larger training sets will be required as the effective dimension increases (and regional structure should be exploited), preliminary experiments on a 12-region prototype under a single stochastic shock indicate that the approach remains numerically stable; a full multi-shock implementation is left for future work.}

It is important to note that IAMs, and the DICE model in particular, have limitations in the model structure and model parameters which are debated in the literature, see e.g. discussions in \cite{pindyck2017use}.
The incorporation of uncertainties into these models is an important improvement.
Our approach demonstrates significant advancements in modeling and solving complex stochastic climate-economy models.
By capturing a wide range of uncertainties and leveraging advanced computational techniques, we contribute to the development of more robust and reliable tools for climate policy analysis.
The continued evolution of these models will be critical in supporting effective and sustainable climate action in the years to come, and the deep least-squares Monte Carlo method provides a useful tool to solve stochastic climate-economy models.

\section*{Acknowledgements}
Aleksandar Arandjelovi{\'c} acknowledges support from the International Cotutelle Macquarie University Research Excellence Scholarship, as well as from the Institute of Statistical Mathematics, Japan.
A significant part of work on this project was carried out while Aleksandar Arandjelovi{\'c} was affiliated with the research unit Financial and Actuarial Mathematics (FAM) at TU Wien in Vienna, Austria as well as the Department of Actuarial Studies and Business Analytics at Macquarie University in Sydney, Australia.
Pavel Shevchenko and Tor Myrvoll acknowledge travel support from the Institute of Statistical Mathematics, Japan.
This work was presented at the international workshop ``Climate Finance \& Risk'' in November 2024 (ISM Japan), the international workshop ``Advances in Risk Modelling and Applications to Finance and Climate Risk'' in July 2024 (WU Wien, Austria), ``Quantitative Methods in Finance'' international conference in December 2025 (Sydney, Australia), and at invited seminars at the University of New South Wales, the Australian National University, Vienna University of Technology and Mahidol University in 2024.
The authors are grateful for the constructive comments received from participants of these events.

\bibliographystyle{plainnat}
\bibliography{bibliography}

\end{document}